\documentclass[aps,prd,12pt]{revtex4}

\begin{document}
\def\erf{\mbox{erf}}
\def\ER{E_{\rm R}}
\def\eth{\epsilon}
\def\emx{\varepsilon}
\def\vesc{v_{\rm esc}}
\def\SD{{\rm SD}}
\def\SI{{\rm SI}}
\renewcommand{\baselinestretch}{1.0}

\title{Direct Search for Dark Matter --- \\
Striking the Balance --- and the Future}

\author{V.A.~Bednyakov}
\affiliation{Dzhelepov Laboratory of Nuclear Problems,
         Joint Institute for Nuclear Research, \\
         141980 Dubna, Russia; E-mail: Vadim.Bednyakov@jinr.ru}
\author{H.V.~Klapdor-Kleingrothaus}
\affiliation{European Center for Scientific Research, Heidelberg, 
Germany\protect\footnote{Postal address of EuCSR Secretariat:
Stahlbergweg 12, 74931 Lobbach, Germany} \protect{\vskip 0.1cm}}

\date{25.~03.~2008}

\begin{abstract}
        Weakly Interacting Massive Particles (WIMPs)
        are among the main candidates for the relic dark matter (DM).
	The idea of the direct DM detection relies on elastic 
	spin-dependent (SD) and spin-independent (SI) 
	interaction of WIMPs with target nuclei. 
        In this review paper the relevant 
        formulae for WIMP event rate calculations are collected.
	For estimations of the WIMP-proton and WIMP-neutron SD
	and SI 	cross sections
	the effective low-energy minimal supersymmetric
	standard model 	is used. 
	The traditional one-coupling-dominance approach for 
	evaluation of the exclusion curves is described.
	Further, the mixed spin-scalar coupling approach is discussed.
	It is demonstrated,  
	taking the high-spin $^{73}$Ge dark matter experiment HDMS
	as an example,
	how one can drastically improve the 
	sensitivity of the exclusion curves within the 
	mixed spin-scalar coupling approach, as well as due to 
	a new procedure of background subtraction from the 
	measured spectrum. 
	A general discussion on the information obtained from 
	exclusion curves is given.
	The necessity of clear WIMP direct detection signatures 
	for a solution of the dark matter problem, is pointed out. 

\vskip 0.3cm 
 
 
\end{abstract}

\maketitle


\section{Introduction}
        To our knowledge the galactic Dark Matter (DM) particles do not emit 
	any detectable amounts of electromagnetic radiation and 
	manifest themselves only gravitationally by 
	affecting other 
	astrophysical objects.
	The first evidence of this kind of substance 
	came from the study of galactic rotation curves,
	i.e. from measurement of the velocity with which 
	stars, globular stellar
	clusters, gas clouds, or dwarf galaxies orbit around their centers
\cite{Zwicky:1933gu}. 
	If the mass of these galaxies was 
	concentrated in their visible parts, the orbital velocity at large
	radii $r$ should decrease 
	in accordance with Kepler's law as $1/\sqrt{r}$.
	Instead, it remains approximately constant to the
	largest radius where it can be measured. 
	This implies that the total mass $M(r)$ felt by an object at a
	radius $r$ must increase linearly with $r$.
	Studies of this type imply that 90\% or more of the mass of
	the large galaxies is in their dark halos
\cite{Freeman:2003aa,Mosher:2007aa,Bertone:2004pz}.

	The mass density averaged over the entire Universe is 
	usually expressed in units of the critical density 
	$\rho_{\rm c} \approx 10^{-29}$g/cm$^3$.
	The dimensionless ratio $\Omega \equiv \rho/\rho_{\rm{c}}= 1$ 
	corresponds to a flat Universe. 
	Analyzes of galactic rotation curves imply $ \Omega \geq 0.1$
(see for example, \cite{Kamionkowski:2007wv,Salucci:2007et,%
Yao:2006px}). 
	Studies of clusters and superclusters of galaxies through 
	gravitational lensing or through measurements of their X-ray 
	temperature, as well as studies of the large-scale streaming of 
	galaxies favor larger values of the total mass density of the 
	Universe $ \Omega \geq 0.3$
(see, for example
\cite{Bergstrom:2000pn,Yao:2006px}).
	Finally, naturalness arguments and most inflationary models prefer 
	$\Omega = 1.0$ to a high accuracy.
	The requirement that the Universe be at least 10 billion years old 
	implies $ \Omega h^2 \leq 1$, where $h $ 
	is the present Hubble parameter in units of 100 km/(sec$\cdot$Mpc)
\cite{Yao:2006px}. 
	The total density of luminous matter only amounts 
	to less than 0.4\% of the critical density
\cite{Kolb:2007gb,Fukugita:2004ee}. 
	Analyzes of Big Bang nucleosynthesis determine the {\em total}\/ 
	baryonic density to lie in the range
	$0.017 \leq \Omega_{\rm b} h^2 \leq 0.024$
\cite{Yao:2006px}. 
	The upper bound implies $\Omega_{\rm b} \leq 0.05,$
	in obvious conflict with the lower bound $ \Omega \geq 0.3$. 
	Most Dark Matter must therefore be non--baryonic.
	Some sort of ``new physics'' is required to describe 
	this exotic matter, 
	{\em beyond}\/ the particles described by the Standard
	Model of particle physics.

\begin{figure}[!ht]
\begin{picture}(50,96)
\put(-43,-37){\includegraphics{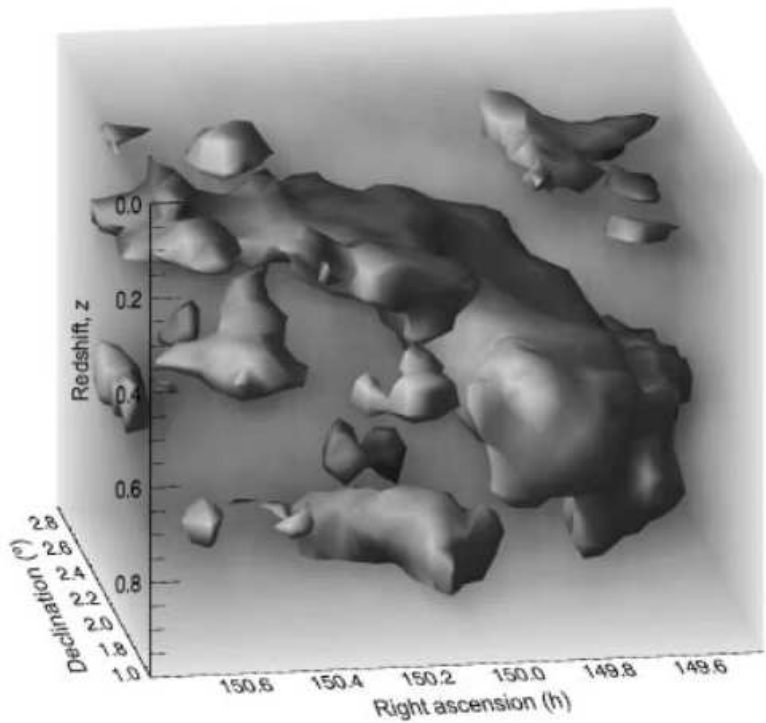}}
\end{picture} 
\caption{Three-dimensional reconstruction of the dark matter distribution. 
    It is obtained by the Hubble Space Telescope Collaboration
\cite{Massey:2007wb}  
    from the differential growth of the gravitational lensing signal
    between many thin discrete redshift slices. 
    The three axes correspond to right ascension, declination and redshift. 
    The distance from the Earth increases toward the bottom
    of the picture.
    For details see the original paper \cite{Massey:2007wb}.
\label{Nature-DM}}
\end{figure}

	Exciting evidence for a flat and 
	accelerating universe was claimed by 
\cite{deBernardis:2000gy,Balbi:2000tg,Linder:2008pp}.
	The position of the first acoustic peak 
	of the angular power spectrum (of the temperature anisotropy of 
	the cosmic microwave background radiation
\cite{Lazkoz:2007ym,Samtleben:2007zz})
 	strongly suggests a flat universe with 
	density parameter $\Omega=1$  
	while the shape of the peak is consistent with the
	density perturbations predicted by models of inflation.
	Data support 	$\Omega = \Omega_{\rm M} + \Omega_{\Lambda}=1$
	where $\Omega_{\rm M}$ is the matter density in the universe and 
	$\Omega_{\Lambda}$ is usually assumed to be a 
	contribution of a non-zero
	cosmological constant (the energy density of the vacuum).
         A first claim for the existence of a non-vanishing 
	 cosmological constant has been made already in 1986 
\cite{Klapdor:1986pg,Klapdor:1986pga}. 
	Recent investigations
        of the Cosmic microwave background temperature anisotropy  
	by the Wilkinson Microwave Anisotropy Probe (WMAP) 
\cite{Spergel:2003cb,Bennett:2003bz,Komatsu:2008hk}	
       and the galaxy power spectrum with the baryon acoustic peak by  
       the Sloan Digital Sky Survey (SDSS) 
\cite{Tegmark:2006az,Percival:2006gt,Tegmark:2005dy}
	supplied us with the 
	values for the cosmological parameters given in 
Table~\ref{WMAP-table}.
\begin{table}[!h]\begin{center}
\begin{tabular}{|l|l|}  \hline
Hubble Constant & ${h = 0.704^{+ 0.015}_{- 0.016}}$ \\
Baryon Density & ${\Omega_{\rm b} h^2 = 0.0219 \pm 0.0007}$ \\
Matter Density & ${\Omega_{\rm M} h^2 = 0.132 \pm 0.004}$\\
\hline
Baryon/Critical Density & ${\Omega_{\rm b} = 0.0442 \pm 0.003}$ \\
Matter/Critical Density & ${\Omega_{\rm M} = 0.249 \pm 0.018}$\\
Total/Critical Density  & $\Omega_{\rm tot} = 1.011 \pm 0.012$\\
Age of the Universe     & ${t_0 = 13.7 \pm 0.2 \mbox{ Gyr}}$ \\
\hline
\end{tabular} 
\label{WMAP-table}
\caption{Some Basic Cosmological Parameters from WMAP and SDSS. From
\cite{Yao:2006px}. 
}
\end{center}\end{table}
        The parameters unambiguously confirm the 
	existence of a large amount of dark matter.
	We omit in this paper discussion of the Dark Energy --- 
	another mysterious substance 
	which is connected with the accelerating Universe and fills 
\cite{Kolb:2007gb}
        the gap between a flat Universe and 
	the measured amount of Dark Matter 
	($\Omega_{\rm DM}+\Omega_{\rm DE}=\Omega_{\rm tot} = 1$).
	In 2006 an exiting ``visualization'' of the invisible dark
	matter substance 
(see Fig.~\ref{Nature-DM})
	has been obtained by means of gravitational lensing 
\cite{Massey:2007wb}. 

	According to the estimates, based on a detailed model of our Galaxy
\cite{Kamionkowski:1997xg},
 	the local density of DM (nearby the solar system) amounts to about
$\rho_{\rm local}^{\rm DM} \simeq 0.3 \ {\rm GeV/cm}^3
\simeq 5 \cdot 10^{-25} {\rm g/cm}^3$,
         with an uncertainty within a factor of two
\cite{Yao:2006px}.
	It is assumed to have a Maxwellian velocity distribution 
	in the galactic rest frame with mean velocity 
	$\bar v \simeq 270$~km/sec
\cite{Freese:1987wu,Lewin:1996rx}. 
	The local flux of DM particles $\chi$ is expected to be 
$\displaystyle 
\Phi_{\rm local}^{\rm DM} \simeq \frac {100 \ {\rm GeV}} {m_\chi}
\cdot 10^5 \ {\rm cm}^{-2} {\rm s}^{-1}.
$
	This value is often considered as a promising basis for 
	direct dark matter search experiments.

	Weakly Interacting Massive Particles (WIMPs) are among the
	most popular candidates for the relic 	dark matter.
	There is no room for such particles in the 
	Standard model of particle physics (SM).
        The lightest supersymmetric (SUSY) particle (LSP), the neutralino
        (being massive, neutral and stable) is currently 
        often assumed to be a favorite WIMP dark matter particle.
        The nuclear recoil energy due to elastic WIMP-nucleus 
        scattering is the quantity to be measured 
        by a terrestrial detector in direct DM detection experiments
\cite{Goodman:1985dc}. 
        Detection of the very rare events of such WIMP interactions 
        is a challenge for modern particle physics, because of
        the very weak WIMP coupling with ordinary matter. 
        The rates expected in SUSY models 
        range from 10 to 10$^{-7}$ 
        events per kilogram detector material and day
(see, for example
\cite{Jungman:1996df,Lewin:1996rx,Ellis:2003eg,Vergados:1996hs,%
Chattopadhyay:2003xi,%
Bednyakov:1994qa,Bednyakov:1996ax,Bednyakov:2000he,%
Bednyakov:2002mb,Bednyakov:1998is,Bednyakov:1997jr}).
         Moreover, for WIMP masses between a few GeV$/c^2$ 
	 and 1 TeV$/c^2$, the energy
         deposited by the recoil nucleus is less than 100 keV.
         Therefore, in order to be able to detect a WIMP, 
         an experiment with a low-energy threshold 
         and an extremely low radioactive background is required. 
         Furthermore, to indeed detect a WIMP one has to 
         unambiguously register some positive signature
         of WIMP-nucleus interactions (directional recoil or
         annual signal modulation)
\cite{Freese:1987wu,Lewin:1996rx}. 
         This means one has to perform a measurement with 
	 a detector of large target mass during several years under
         extremely low radioactive background conditions
         (see also the discussions of other complications in 
\cite{Gondolo:2005qp,Bertone:2007ki,Hooper:2008sn}). 
         Despite of all these problems huge effort is at present 
	 put into direct detection of DM particles
(see for example, 
\cite{Yao:2006px,Akerib:2006ks,Freeman:2003aa,Akimov:2001nr}). 

          Till now only the DAMA (DArk MAtter) collaboration claims 
\cite{Bernabei:2000qi,Bernabei:2001ve,Bernabei:2003za,Bernabei:2003wy}
        observation of first evidence for a dark matter signal
        due to registration of the predicted annual
        modulation of specific shape and amplitude 
        due to the combined motions of the Earth and Sun 
        around the galactic center
\cite{Freese:1987wu}.
        Aimed since more than one decade at the direct detection
        of DM particles, 
        the DAMA experiment (DAMA/NaI) with 100 kg
        of highly radio-pure NaI(Tl) scintillator detectors  
        successfully operated till July 2002 
        at the Gran Sasso National Laboratory (LNGS) of the I.N.F.N.
	On the basis of the results obtained over 7 annual cycles 
	(107731~kg$\cdot$day total DAMA exposure)
	the presence of a WIMP-model independent 
	annual modulation signature was observed at a 6.3 $\sigma$ C.L. 
\cite{Bernabei:2003za}.
	The main result of the DAMA observation 
	of the annual modulation signature is a low-mass region of the WIMPs
	($40 < m_\chi < 150$~GeV) 
	and relatively high allowed SI or/and SD cross sections
	(for example, 
	$ 1\cdot10^{-7}~{\rm pb}<\sigma^p_\SI(0)<3\cdot 10^{-5}~{\rm pb}$), 
	provided these WIMPs are cold dark matter particles. 

        Although there are other experiments like
        EDELWEISS, CDMS, etc, which give sensitive exclusion curves,
        no one of them at present has the sensitivity to look 
        for the modulation effect.
        Due to the small target masses and short running times 
	these experiments 	are unable to see a positive 
        annual modulation signature of the WIMP interactions.
        Some other experiments
        with much larger mass targets (mostly NaI) 
        unfortunately are also unable to 
        register the positive signature due to 
        not good enough background conditions
(see for example, \cite{Alner:2005kt,Cebrian:2002vd,Yoshida:2000df}).
  	Often the results of these and the DAMA experiment 
  	have been compared not on the basis of a complete 
  	analysis including simultaneously SI and SD WIMP 
	nucleus interaction. 
	This sometimes gives rise to quite some confusion in the 
	literature 
	(for a discussion see 
\cite{Bednyakov:2004be,Bednyakov:2005qp}), and to 
	attempts to reconcile an artificial  
        DAMA ``conflict'' with the other experiments
\cite{Copi:2002hm,Kurylov:2003ra,Tucker-Smith:2004jv,Gelmini:2004gm,%
Savage:2004fn,Gondolo:2005hh,Gelmini:2005fb}.

	Despite of the well-known attempts of the DAMA collaboration 
	to prove this observation with 
	a new larger NaI setup DAMA/LIBRA
\cite{Bernabei:2006tz}, 
	it is obvious that such a serious claim
	should be verified 
	at least by one other completely independent experiment.  
	To confirm this DAMA result one should perform
	a new experiment which would have (in reasonable time) 
	the same or better sensitivity to the annual modulation signal
	(and also it would be reasonable  to locate this new 
	setup in another low-background underground laboratory).
	This mission, in particular, 
	could be executed by new-generation experiments
	with large enough mass of germanium 
	high purity (HP) detectors both 
	with spin ($^{73}$Ge) and spin-less (natural Ge). 
	 Despite of obviously necessary strong figthing against backgrounds, 
         the main direction in development of new-generation 
         DM detectors  concerns remarkable enlargement of the
         target mass to be able to observe these
         positive signatures, and thus to detect DM and
         to prove, or disprove the DAMA claim.
        In particular, an enlarged version of the 
	EDELWEISS setup with 40 kg bolometric Ge detectors 
\cite{Sanglard:2006hd} together with, perhaps, SuperCDMS 
\cite{Akerib:2006rr,Brink:2005ej}, 
         as well as enlarged ZEPLIN 
\cite{Akimov:2006qw} 
        or KIMS 
\cite{Lee.:2007qn}  experiments might become 
	sensitive to the annual modulation in some future.

	The main efforts (and expectations) in present direct dark 
	matter searches are concentrated in the field of so-called  
        spin-independent (or scalar) interaction of a dark matter 
	WIMP with a target nucleus. 
	This is because it was 	found theoretically that
	for heavy enough nuclei
	this spin-independent (SI) interaction of 
        DM particles with nuclei usually gives the 
	dominant contribution to the expected event rate of its detection.
        The reason is the strong (proportional to the squared mass 
        of the target nucleus) 
        enhancement of the SI WIMP-nucleus interaction. 

	The spin-1/2 WIMP particles, like the LSP neutralinos,
        interact with ordinary matter predominantly
        by means of axial vector (spin-dependent) and
        vector (spin-independent) couplings.
        There is some revival of interest in the 
	WIMP-nucleus spin-dependent interaction from both theoretical 
    (see e.g. 
\cite{Engel:1991wq,Bottino:2003cz,%
Bednyakov:1994te,Bednyakov:2000he,Bednyakov:2002mb,Bednyakov:2003wf,%
Bednyakov:2004xq,Bednyakov:2005qp})
     and experimental (see e.g. 
\cite{Girard:2005pt,Girard:2005dq,%
Giuliani:2004uk,Giuliani:2005bd,%
Savage:2004fn,Benoit:2004tt,Tanimori:2003xs,%
Moulin:2005sx,Mayet:2002ke,Klapdor-Kleingrothaus:2005rn})
      points of view.  
     There are some proposals aimed at direct DM detection with 
     relatively low-mass isotope targets 
\cite{Girard:2005pt,Girard:2005dq,Tanimori:2003xs,Ovchinnikov:2003AA,%
Moulin:2005sx,Mayet:2002ke}
      as well as some attempts to design and construct 
      a DM detector which is sensitive to the nuclear recoil direction
\cite{Alner:2004cw,Snowden-Ifft:1999hz,Gaitskell:1996cv,Sekiya:2004ma,%
Morgan:2004ys,Vergados:2000cp,Vergados:2002bb}.
       Low-mass targets 
      make preference for the low-mass WIMPs 
      and are more sensitive to the spin-dependent 
      WIMP-nucleus interaction as well 
\cite{Jungman:1996df,Engel:1991wq,Divari:2000dc,%
Bednyakov:1994te,Bednyakov:2000he,Bednyakov:2004xq,Bednyakov:1997ax}.

        There are at least three reasons to think that 
	SD (or axial-vector) interaction of the DM WIMPs with nuclei
        could be very important. 
        First, contrary to the only one constraint for SUSY models available 
        from the scalar WIMP-nucleus interaction, the spin WIMP-nucleus 
        interaction supplies us with two such constraints (see for example 
\cite{Bednyakov:1994te} and formulae below).
        Second, one can notice 
\cite{Bednyakov:2000he,Bednyakov:2002mb}
        that even with a very sensitive DM detector
        (say, with a sensitivity of $10^{-5}\,$events$/$day$/$kg)
        which is sensitive only to the WIMP-nucleus 
        scalar interaction (with spin-less target nuclei) 
        one can, in principle, miss a DM signal. 
        To safely avoid such a situation one should
        have a spin-sensitive DM detector, i.e. a detector 
        with spin-non-zero target nuclei.
        Finally, there is a complicated 
        nuclear spin structure, which 
	possesses the so-called long $q$-tail form-factor behavior. 
        The SI WIMP-nucleus cross section, despite being proportional 
	to $A^2$, vanishes very quickly (exponentially) 
	with increasing momentum transfer $q^2$. 
	The SD WIMP-nucleus cross section decreases not so
	quickly with $q^2$ and remains 
 	still final at the recoil energies
	($E_{\rm R} = q^2 /(2 M_A )$),
	where the SI cross section is already zero.
        Therefore for heavy mass target nuclei and heavy WIMP masses
	the SD efficiency to detect a DM signal 
	could be much higher than the SI efficiency
\cite{Engel:1991wq}.  
	Therefore, simultaneous study of both 
        spin-dependent and spin-independent interactions of the
        DM particles with nuclei significantly increases 
	the chance to observe the DM signal
\cite{Bednyakov:2007zz,Bednyakov:2003wf,Bednyakov:2002mb,Bednyakov:2004be}. 

          Following R.Bernabei et al.
\cite{Bernabei:2003za,Bernabei:2001ve}  it was stressed in 
\cite{Bednyakov:2004be,Bednyakov:2005qp}
          that for analyzing data from 
          DM detectors with spin-non-zero targets 
          one should use the so-called mixed
          spin-scalar coupling approach.
         This approach is used to demonstrate, taking  
	 the high-spin $^{73}$Ge detector HDMS
\cite{Klapdor-Kleingrothaus:2002pg,Klapdor-Kleingrothaus:2000uh}
         as example, how one can stronger
	 improve the exclusion curves. 
         The mixed spin-scalar coupling approach allowed one
          to extract information about both 
          SI and SD WIMP-nucleon cross sections 
          analyzing background spectra from the two HDMS setups
          (prototype and final) simultaneously. 
          This procedure allows an improvement 
(see our new analysis in 
\cite{Bednyakov:2007yf})
          of the exclusion curves relative to the relevant curves 
          obtained in the traditional  
	  one-coupling dominance approach for the HDMS in 
\cite{Klapdor-Kleingrothaus:2005rn}. 

\smallskip
        The present paper is organized as follows.
        In the next Section the main formulae for 
	event rate calculations are collected.
	In Section III 
	the effective low-energy minimal supersymmetric
	standard model (effMSSM) is used for calculation of 
	the WIMP-proton and WIMP-neutron SD and SI cross sections. 
	In Section IV the traditional one coupling dominance approach for 
	evaluation of the exclusion curves is discussed.
	In Section 
	V the mixed spin-scalar couplings approach is described, 
	the DAMA-inspired exclusion domains for both above-mentioned 
	couplings are given and compared with SUSY calculations.
	In Section 
	VI the mixed spin-scalar coupling scheme is applied to 
	the high-spin $^{73}$Ge dark matter search experiment HDMS. 
	It is demonstated how one can strongly 
	improve the 
	quality of the exclusion curves within the 
	mixed spin-scalar coupling approach as well as by using 
	a new procedure of background subtraction from the 
	measured spectrum. 
	In Section 
	VII a general discussion is given.
	The conclusion summarizes the main items of this review paper.

\section{Event rate and cross sections}
\begin{figure}[t!] 
\begin{picture}(100,87) 
\put(-2,-6){\includegraphics{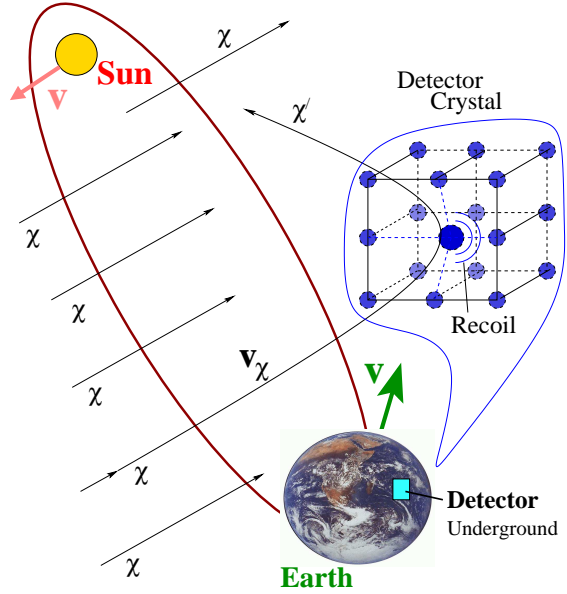}}
\end{picture} 
\caption{Detection of the cold dark matter (WIMPs) by elastic 
  scattering from target nuclei in the detector.
  Due to the expected annual modulation signature of the event rate
  Eq.~(\ref{Definitions.diff.rate}) the Sun-Earth system 
  is a particularly proper setup for successful direct DM detection.
\label{DDDDD}}
\end{figure}

        Many experiments try to detect directly a relic DM WIMP 
	(or neutralino) $\chi$ 	with mass ${m_\chi}$ 
	via its elastic scattering on a target nucleus $(A,Z)$. 
	The nuclear recoil energy $E_{\rm R}$ 
	is measured by a proper detector deeply underground 
(Fig.~\ref{DDDDD}).
	The differential event rate in respect to the recoil 
	energy (the spectrum) is the subject of the measurements.
	The rate depends on the density and the velocity distribution of
        the relic WIMPs in the solar vicinity $f(v)$ and
        the cross section of WIMP-nucleus elastic scattering
\cite{Jungman:1996df,Lewin:1996rx,Smith:1990kw,Bednyakov:1999yr,Bednyakov:1996yt,Bednyakov:1997ax,Bednyakov:1997jr,Bednyakov:1994qa}.
	The differential event rate per unit mass of 
	the target material has the form
\begin{equation}
\label{Definitions.diff.rate}
	\frac{dR}{dE_{\rm R}} = N_T \frac{\rho_\chi}{m_\chi} 
	\int^{v_{\max}}_{v_{\min}} dv f(v) v
	{\frac{d\sigma^A}{dq^2}} (v, q^2). 
\end{equation}
        We assume these WIMPs (or neutralinos) 
	to be the dominant component of
        the DM halo of our Galaxy with a density
        $\rho_{\chi}$ = 0.3 GeV$/$cm$^{3}$ in the solar vicinity.
        The (real) nuclear recoil energy
	$E_{\rm R} = q^2 /(2 M_A )$ is typically about $10^{-6} m_{\chi}$ and 
	$N_T$ is the number density of target nuclei with mass $M_A$. 
	$v_{\max} = v_{\rm esc} \approx 600$~km/s, \ 
	$v_{\min}=\left(M_A E_{\rm R}/2 \mu_{A}^2\right)^{1/2}$ is
	the minimal WIMP velocity which still can 
	produce the recoil energy $E_{\rm R}$.
	The WIMP-nucleus differential 
	elastic scattering cross section 
	for spin-non-zero ($J\neq 0$) 
	nuclei contains coherent (spin-independent, or
	SI) and axial (spin-dependent, or SD) terms
\cite{Engel:1992bf,Engel:1991wq,Ressell:1993qm}: 
\begin{eqnarray} \label{drateEPV}
\frac{d\sigma^A}{dq^2}(v,q^2)
&=&  \frac{S^A_{\rm SD} (q^2)}{v^2 (2J+1)} 
    +\frac{S^A_{\rm SI} (q^2)}{v^2 (2J+1)}
\label{Definitions.cross.section}
= \frac{\sigma^A_{\rm SD}(0)}{4\mu_A^2 v^2}F^2_{\rm SD}(q^2)
           +\frac{\sigma^A_{\rm SI}(0)}{4\mu_A^2 v^2}F^2_{\rm SI}(q^2).
\label{eq:2}
\end{eqnarray}
 	The normalized ($F^2_{\rm SD,SI}(0) = 1$)
	finite-momentum-transfer nuclear form-factors
$\displaystyle
F^2_{\rm SD,SI}(q^2) = \frac{S^{A}_{\rm SD,SI}(q^2)}{S^{A}_{\rm SD,SI}(0)} 
$
	can be expressed through the nuclear structure functions as follows
\cite{Engel:1992bf,Engel:1991wq,Ressell:1993qm}: 
\begin{eqnarray}
\label{Definitions.scalar.structure.function}
S^{A}_{\rm SI}(q) 
	&=& 
	\sum_{L\, {\rm even}} 
        \vert\langle J \vert\vert {\cal C}_L(q) \vert\vert J \rangle \vert^2 
	\simeq  
	\vert\langle J \vert\vert {\cal C}_0(q) \vert\vert J \rangle \vert^2 ,
\nonumber \\ 
S^A_{\rm SD}(q) 
	&=& 
	\sum_{L\, {\rm odd}} \big( 
	\vert\langle N \vert\vert {\cal T}^{el5}_L(q) 
	\vert\vert N \rangle\vert^2 + \vert\langle N \vert\vert 
	{\cal L}^5_L (q) \vert\vert N \rangle\vert^2\big). 
\label{Definitions.spin.structure.function}
\label{SF-definition}
\label{eq:4}
\end{eqnarray} 
	The explicit form of the transverse electric ${\cal T}^{el5}(q)$ 
	and longitudinal ${\cal L}^5(q)$ multipole projections of the
	axial vector current operator and the scalar function ${\cal C}_L(q)$ 
	can be found in 
\cite{Engel:1992bf,Engel:1991wq,Ressell:1993qm,%
Bednyakov:2004xq,Bednyakov:2006ux}.
	For 	$q=0$ the nuclear SD and SI cross sections  
	can be presented as follows
\begin{eqnarray}
\label{Definitions.scalar.zero.momentum}
\sigma^A_{\rm SI}(0) 
        &=& \frac{4\mu_A^2 \ S^{}_{\rm SI}(0)}{(2J+1)}\! =\!
             \frac{\mu_A^2}{\mu^2_p}A^2 \sigma^{p}_{{\rm SI}}(0), \\ 
\label{Definitions.spin.zero.momentum}
\sigma^A_{\rm SD}(0)
        &=&  \frac{4\mu_A^2 S^{}_{\rm SD}(0)}{(2J+1)}\! =\!
             \frac{4\mu_A^2}{\pi}\frac{(J+1)}{J}
             \left\{a_p\langle {\bf S}^A_p\rangle 
                  + a_n\langle {\bf S}^A_n\rangle\right\}^2\\
\label{Definitions.spin.zero.momentum.Tovei}
&=&
	\frac{\mu_A^2}{\mu_p^2}
	\frac43 \frac{J+1}{J}
	\sigma^{pn}_{\rm SD}(0)	
	\left\{ \langle {\bf S}^A_p\rangle \cos\theta
               +\langle {\bf S}^A_n\rangle \sin\theta   
       \right\}^2.
\label{Definitions.spin.zero.momentum.Bernabei}
\end{eqnarray} 
	Following Bernabei et al.
\cite{Bernabei:2003za,Bernabei:2001ve}
	the effective spin WIMP-nucleon cross section
	$\sigma^{pn}_{\rm SD}(0)$
	and the coupling mixing angle $\theta$ were introduced
\begin{eqnarray}
\label{effectiveSD-cs}
\sigma^{pn}_{\rm SD}(0)
	&=& \frac{\mu_p^2}{\pi}\frac43 
		\Bigl[ a_p^2 +a_n^2 \Bigr], \qquad
\tan\theta = \frac{{a}_{n}}{{a}_{p}}; \\
\label{effectiveSD-cs-pn} 
\sigma^p_{\rm SD}&=&\sigma^{pn}_{\rm SD} \cdot \cos^2 \theta, \quad
\sigma^n_{\rm SD}=\sigma^{pn}_{\rm SD} \cdot \sin^2 \theta.
\end{eqnarray}
	Here, $\displaystyle \mu_A = \frac{m_\chi M_A}{m_\chi+ M_A}$
	is the reduced mass of the neutralino and the nucleus, 
	and it is assumed that $\mu^2_{n}=\mu^2_{p}$.
	The dependence on effective WIMP-quark 
	(in SUSY neutralino-quark) couplings 
	${\cal C}_{q}$ and ${\cal A}_{q}$ in the underlying theory 
\begin{equation}
{\cal  L}_{\rm eff} = \sum_{q}^{}\left( 
	{\cal A}_{q}\cdot
      \bar\chi\gamma_\mu\gamma_5\chi\cdot
                \bar q\gamma^\mu\gamma_5 q + 
	{\cal C}_{q}\cdot\bar\chi\chi\cdot\bar q q
	\right)      \ + ... 
\label{Definitions.effective.lagrangian}
\end{equation}
	and on the spin ($\Delta^{(p,n)}_q$)
	and the mass or scalar ($f^{(p)}_q \approx f^{(n)}_q$) 
	structure of the proton and neutron 
	enter into these formulae via the zero-momentum-transfer 
	WIMP-proton and WIMP-neutron SI and SD cross sections: 
\begin{eqnarray}
\label{Definitions.scalar.zero.cs}
\label{SI-cs-Deff.at-zero}
\sigma^{p}_{{\rm SI}}(0) 
	= 4 \frac{\mu_p^2}{\pi}c_{0}^2,
&\qquad&
	c^{}_0 = c^{p,n}_0 = \sum_q {\cal C}_{q} f^{(p,n)}_q; \\
\sigma^{p,n}_{{\rm SD}}(0)  
 	=  12 \frac{\mu_{p,n}^2}{\pi}{a}^2_{p,n} 
&\qquad&
	a_p =\sum_q {\cal A}_{q} \Delta^{(p)}_q, \quad 
	a_n =\sum_q {\cal A}_{q} \Delta^{(n)}_q.
\label{SD-cs-Deff.at-zero}
\end{eqnarray}
	The factors $\Delta_{q}^{(p,n)}$, which parameterize the quark 
	spin content of the nucleon, are defined as
	$ \displaystyle 2 \Delta_q^{(n,p)} s^\mu  \equiv 
          \langle p,s| \bar{\psi}_q\gamma^\mu \gamma_5 \psi_q    
          |p,s \rangle_{(p,n)}$.
	The quantity $\langle {\bf S}^A_{p(n)} \rangle $ denotes 
	the total spin of protons 
	(neutrons) averaged over all $A$ nucleons of the nucleus $(A,Z)$:
\begin{equation}
\langle {\bf S}^A_{p(n)} \rangle 
     \equiv \langle A \vert  {\bf S}^A_{p(n)} \vert A \rangle 
     = \langle A \vert  \sum_i^A {\bf s}^i_{p(n)} \vert A \rangle. 
\end{equation}
        The mean velocity $\langle v \rangle$ of 
	the relic neutralinos of our Galaxy
	is about  $ 300~{\rm km/s} = 10^{-3} c$.
	Assuming 
        $q_{\rm max}R \ll 1$, where $R$ is the nuclear radius 
	and $q_{\rm max} = 2 \mu_A v$ is the maximum of the momentum 
	transfer in the process of the $\chi A$ scattering, the 
        spin-dependent matrix element  takes a simple form
({\em zero momentum transfer limit})\
\cite{Engel:1995gw,Ressell:1997kx}:
\begin{equation}
\label{Definitions.matrix.element}
 {\cal M} = C \langle A\vert a_p {\bf S}_p + a_n {\bf S}_n
 	\vert A \rangle \cdot {\bf s}_{\chi}
 	  = C \Lambda \langle A\vert {\bf J}
	 \vert A \rangle \cdot {\bf s}_{\chi}.
\label{eq:10}
\end{equation}
	Here, ${\bf s}_{\chi}$ denotes the spin of the neutralino, and 
\begin{equation}
 \Lambda = {{\langle N\vert a_p {\bf S}_p + a_n {\bf S}_n
\vert N \rangle}\over{\langle N\vert {\bf J}
\vert N \rangle}} =
{{\langle N\vert ( a_p {\bf S}_p + a_n {\bf S}_n ) \cdot {\bf J}
\vert N \rangle}\over{ J(J+1)
}} = 
{{a_p \langle {\bf S}_p \rangle}\over{J}}  +
{{a_n \langle {\bf S}_n \rangle}\over{J}}.
\label{eq:11}
\end{equation}
        Note a coupling  of the spin of $\chi$ to the spin carried
	by the protons and the neutrons.  The uncertainties arising from 
        the electroweak and QCD scale physics are incorporated 
        in the factors $a_p$ and $a_n$. 
	The normalization factor $C$ involves the coupling
	constants, the  masses of the exchanged bosons and 
        the mixing parameters relevant to the 
        lightest supersymmetric particle (LSP), i.e., it is not
        related to the associated nuclear matrix elements
\cite{Griest:1988ma}.  
	In the limit of zero momentum transfer $q=0$ 
	the spin structure function in Eq. 
(\ref{Definitions.spin.structure.function}) reduces to the form
\begin{eqnarray*}
S^A(0)={1\over{4 \pi}} \vert\langle A \vert\vert\sum_i
    {1\over{2}}(a_0 + a_1 \tau_3^i) {\bf \sigma}_i\vert\vert A \rangle\vert^2 
     =\frac{2J+1}{\pi} J(J+1) \Lambda^2. 
\end{eqnarray*}
        For the most interesting
	isotopes either $\langle{\bf S}^A_{p}\rangle$ 
	or $\langle{\bf S}^A_{n}\rangle$ dominates
	($\langle{\bf S}^A_{n(p)}\rangle \ll \langle{\bf S}^A_{p(n)}\rangle$).
	See, for example, 
Table~\ref{Nuclear.spin.main.table}.

\begin{table}[t!] 
{
\caption{Zero momentum spin structure of nuclei in different models. 
	The measured magnetic moments used as input 
	are enclosed in parentheses. 
        The variation of the $\langle {\bf S}^A_{p}\rangle$
        and $\langle {\bf S}^A_{n}\rangle$ for fixed $A$ 
	reflects the level of  inaccuracy and complexity 
        of the current nuclear structure calculations.
	From 
\cite{Bednyakov:2004xq}.
\label{Nuclear.spin.main.table}}
\label{Nuclear.spin.main.table.71-95}
\begin{center}
\vspace*{-7pt}
\begin{tabular}{lrrr}
\hline
\hline
$^{73}$Ge~($L_J=G_{9/2}$) & ~~~~~~~~$\langle {\bf S}_p \rangle$ & 
~~~~~~~~$\langle {\bf S}_n \rangle$ & ~~~~~~~~$\mu$ (in $\mu_N$) \\ \hline
ISPSM, Ellis--Flores~\cite{Ellis:1988sh,Ellis:1991ef}
	&    0	  & $0.5$		& $-1.913$ \\ 
OGM, Engel--Vogel~\cite{Engel:1989ix} 	
	&    0	  & $0.23$ 	&$(-0.879)_{\rm exp}$ \\ 
IBFM, Iachello et al.~\cite{Iachello:1991ut} and \cite{Ressell:1993qm}
	&$-0.009$ & $0.469$ &$-1.785$\\ 
IBFM (quenched), 
	Iachello et al.~\cite{Iachello:1991ut} and \cite{Ressell:1993qm}
	&$-0.005$  & $0.245$ &$(-0.879)_{\rm exp}$ \\
TFFS, Nikolaev--Klapdor-Kleingrothaus, \cite{Nikolaev:1993dd} 
	&$0$   & $0.34$ & --- \\ 
SM (small), Ressell et al.~\cite{Ressell:1993qm} 
	&$0.005$   & $0.496$ &$-1.468$ \\ 
SM (large), Ressell et al.~\cite{Ressell:1993qm} 
	&$0.011$   & $0.468$ &$-1.239$ \\ 
SM (large, quenched), Ressell et al.~\cite{Ressell:1993qm} 
	&$0.009$   & $0.372$ &$(-0.879)_{\rm exp}$ \\ 
``Hybrid'' SM, Dimitrov et al.~\cite{Dimitrov:1995gc}           
	& $0.030$ & $0.378$ & $-0.920$ \\ 
\hline
$^{127}$I~($L_J=D_{5/2}$) & 
~~~~~~~~$\langle {\bf S}_p \rangle$ & ~~~~~~~~$\langle {\bf S}_n \rangle$ & 
~~~~~~~~$\mu$ (in $\mu_N$) \\ \hline
ISPSM, Ellis--Flores~\cite{Ellis:1991ef,Ellis:1993vh}
	&$1/2$ & 0 &  $4.793$ \\ 
OGM, Engel--Vogel~\cite{Engel:1989ix} 
	&$0.07$ & 0 & $(2.813)_{\rm exp}$ \\ 
IBFM, Iachello et al.~\cite{Iachello:1991ut}
	&$0.464$ & $0.010$ & $(2.813)_{\rm exp}$ \\ 
IBFM (quenched), 
	Iachello et al.~\cite{Iachello:1991ut}
	&$0.154$  & $0.003$ & $(2.813)_{\rm exp}$ \\ 
TFFS, Nikolaev--Klapdor-Kleingrothaus, \cite{Nikolaev:1993dd} 
	& $0.15$ & 0 & --- \\ 
SM (Bonn A), Ressell--Dean~\cite{Ressell:1997kx} 
	& $0.309$ & $0.075$ & $2.775~\{2.470\}_{\rm eff}$  \\ 
SM (Nijmegen II), Ressell--Dean~\cite{Ressell:1997kx} 
	& $0.354$ & $0.064$ & $3.150~\{2.7930\}_{\rm eff}$ \\ 
\hline
$^{131}$Xe~($L_J=D_{3/2}$) & 
~~~~~~~~$\langle {\bf S}_p \rangle$ & 
~~~~~~~~$\langle {\bf S}_n \rangle$ & 
~~~~~~~~$\mu$ (in $\mu_N$) \\ \hline 
ISPSM, Ellis--Flores~\cite{Ellis:1988sh,Ellis:1991ef}
	& 0 & $-0.3$ & $1.148$ \\
OGM, Engel--Vogel~\cite{Engel:1989ix} 
	& 0.0  & $-0.18$ 	& $(0.692)_{\rm exp}$ \\ 
IBFM, Iachello et al.~\cite{Iachello:1991ut}
	&$0.000$ & $-0.280$ 	& $(0.692)_{\rm exp}$ \\
IBFM (quenched), Iachello et al.~\cite{Iachello:1991ut}
	&$0.000$  & $-0.168$ 	& $(0.692)_{\rm exp}$ \\
TFFS, Nikolaev--Klapdor-Kleingrothaus, \cite{Nikolaev:1993dd} 
	&   & $-0.186$ & --- \\ 
SM (Bonn A), Ressell--Dean~\cite{Ressell:1997kx} 
	& $-0.009$ & $-0.227$ & $0.980~\{0.637\}_{\rm eff}$  \\ 
SM (Nijmegen II), Ressell--Dean~\cite{Ressell:1997kx} 
	& $-0.012$ & $-0.217$ & $0.979~\{0.347\}_{\rm eff}$ \\  
QTDA, Engel~\cite{Engel:1991wq} 
	& $-0.041$ & $-0.236$ & 0.70 \\  
\hline
\end{tabular} \end{center}
}\end{table} 
   
        The differential event rate 
(\ref{Definitions.diff.rate}) can be given also in the form 
\cite{Bernabei:2003za,Bednyakov:2004be}: 
\begin{eqnarray}
\label{Definitions.diff.rate1}
\frac{dR(\ER)}{d\ER} 
	&=& \kappa^{}_\SI(\ER,m_\chi)\,\sigma_\SI
         +\kappa^{}_\SD(\ER,m_\chi)\,\sigma_\SD. \\
\nonumber
\kappa^{}_\SI(\ER,m_\chi)
&=&       N_T \frac{\rho_\chi M_A}{2 m_\chi \mu_p^2 } 
          B_\SI(\ER) \left[ M_A^2 \right],\\
\kappa^{}_\SD(\ER,m_\chi)
&=& 
\label{structure} 
     N_T \frac{\rho_\chi M_A}{2 m_\chi \mu_p^2 } B_\SD(\ER) 
        \left[\frac43 \frac{J+1}{J}
        \left(\langle {\bf S}_p \rangle \cos\theta
            + \langle {\bf S}_n \rangle \sin\theta   
       \right)^2\right]  ,\\
B_{\SI,\SD}(\ER) 
&=& \nonumber
        \frac{\langle v \rangle}{\langle v^2 \rangle}
        F^2_{\SI,\SD}(\ER)I(\ER). 
\end{eqnarray}
        The dimensionless integral $I(\ER)$ 
        is a dark-matter-particle velocity distribution 
	correction (see Eq.
(\ref{Definitions.diff.rate1})):
\begin{equation}\label{I_ER}
I(\ER)= \frac{ \langle v^2 \rangle}{ \langle v \rangle }
 \int_{x_{\min}} \frac{f(x)}{v} dx 
    = \frac{\sqrt{\pi}}{2}
\frac{3 + 2 \eta^2}{{\sqrt{\pi}}(1+2\eta^2)\erf(\eta) + 2\eta e^{-\eta^2}}
                [\erf(x_{\min}+\eta) - \erf(x_{\min}-\eta)],
\end{equation}
        where one assumes that in the rest frame of our Galaxy 
        WIMPs have a Maxwell-Boltzmann velocity distribution, and
        uses the dimensionless Earth speed with respect to the halo 
	$\eta$,\ as well as 
        $\displaystyle x_{\min}^2 = 
        \frac{3}{4}\frac{M_A\ER}{\mu^2_A{\bar{v}}^2}$
\cite{Freese:1987wu,Lewin:1996rx}. 
        The error function is 
        $\displaystyle \erf(x) = \frac{2}{\sqrt{\pi}}\int_0^x dt e^{-t^2}$.
        The velocity variable is the dispersion $\bar{v}\simeq 270\,$km$/$c.
        The mean WIMP velocity 
        ${\langle v \rangle} = \sqrt{\frac{5}{3}} \bar{v}$.
        We also assume both form-factors 
        $F^2_{\SI,\SD}(\ER)$ in the simplest Gaussian form following
\cite{Ellis:1988sh,Ellis:1991ef}.
         In particular, this allows rather simple formulae
(see Eq.~(\ref{structure})) to be used. 
	Integrating the differential rate 
Eq.~(\ref{Definitions.diff.rate}) 
	from the recoil energy threshold $\eth$ 
	to some maximal energy $\emx$
        one obtains the total detection rate $R(\eth, \emx)$
	as a sum of the SD and SI terms:
\begin{eqnarray}\label{Definitions.total.rate}
\label{for-toy-mixig}
R(\eth, \emx)&=&
     R_{\SI}(\eth, \emx) + R_{\SD}(\eth, \emx) = 
      \int^{\emx}_{\eth} d\ER \kappa^{}_\SI(\ER,m_\chi)\,\sigma_\SI
    + \int^{\emx}_{\eth} d\ER \kappa^{}_\SD(\ER,m_\chi)\,\sigma_\SD.
\end{eqnarray}
	To accurately estimate the event rate $R(\eth, \emx)$
	one needs to know a number of quite uncertain 
	astrophysical and nuclear structure parameters
	as well as the very specific characteristics of an experimental setup
	(see, for example, discussions in 
\cite{Bernabei:2003xg,Bernabei:2003za}). 
 
	As $m_{\chi}$ increases, 
	the product $qR$ starts to become non-negligible 
	and {\em the finite momentum transfer limit}\/
	must be considered
\cite{Engel:1992bf,Ressell:1993qm,Ressell:1997kx,%
Bednyakov:2004xq,Bednyakov:2006ux}. 
	The formalism is a straightforward extension of that
	developed for the study of weak and electromagnetic 
	semi-leptonic interactions in nuclei 
\cite{Ressell:1997kx,Ressell:1993qm}.
	With the isoscalar spin coupling constant $a_0 = a_n + a_p$
	and the isovector spin coupling constant
	$a_1 = a_p - a_n$ one can split 
	the nuclear structure function $S^A_{}(q)$ 
(from Eqs. (\ref{Definitions.cross.section}) and 
     (\ref{Definitions.scalar.structure.function}))
	into a pure
	isoscalar term, $S^A_{00}(q)$, a pure isovector term, $S^A_{11}(q)$, 
	and an interference term, $S^A_{01}(q)$, in the following way:
\begin{equation}
\label{Definitions.spin.decomposition}
S^A_{}(q) = a_0^2 S^A_{00}(q) + a_1^2 S^A_{11}(q) + a_0 a_1 S^A_{01}(q).
\end{equation}
	The relations 
$
S^A_{00}(0) = C(J)(\langle {\bf S}_p \rangle + \langle {\bf S}_n \rangle)^2,
$ 
$S^A_{11}(0) = C(J)(\langle {\bf S}_p \rangle - \langle {\bf S}_n \rangle)^2,
$ and 
$S^A_{01}(0) =2C(J)(\langle {\bf S}^2_p \rangle - \langle {\bf S}^2_n \rangle)
$ with
$\displaystyle
C(J)= \frac{2J+1}{4\pi}\frac{J+1}{J},
$
       connect the nuclear spin structure function $S^A(q=0)$ 
       with proton $\langle {\bf S}_p \rangle$ 
       and neutron $\langle {\bf S}_n \rangle$
       spin contributions averaged over the nucleus
\cite{Bednyakov:2006ux}. 

       These three partial structure  functions $S^A_{ij}(q)$ 
       allow calculation 
       of spin-dependent cross sections for any heavy Majorana particle
       as well as for the neutralino with arbitrary composition
\cite{Engel:1995gw}.

        The first model to estimate the spin content in the nucleus
        for the dark matter search was the 
        independent single-particle shell model ({ISPSM})
        used originally by Goodman and Witten 
\cite{Goodman:1985dc} and later in
\cite{Drukier:1986tm,Ellis:1988sh,Smith:1990kw}.
        There are several approaches 
	to more accurate calculations of the nuclear
        structure effects relevant to the dark matter detection.
	The list of the models includes the
        Odd Group Model ({OGM}) of Engel and Vogel
\cite{Engel:1989ix} and their extended OGM ({EOGM})
\cite{Engel:1989ix,Engel:1992bf}; 
        Interacting Boson Fermion Model ({IBFM}) of
        Iachello, Krauss, and Maino      
\cite{Iachello:1991ut};
        Theory of Finite Fermi Systems ({TFFS}) of 
        Nikolaev and Klapdor-Kleingrothaus
\cite{Nikolaev:1993dd};
        Quasi Tamm-Dancoff Approximation ({QTDA}) of Engel
\cite{Engel:1991wq};
        different shell model treatments ({SM}) by Pacheco and Strottman 
\cite{Pacheco:1989jz};
        by Engel, Pittel, Ormand and Vogel 
\cite{Engel:1992qb} and 
        Engel, Ressell, Towner and Ormand,
\cite{Engel:1995gw},    
        by Ressell et al.
\cite{Ressell:1993qm} and 
        Ressell and Dean
\cite{Ressell:1997kx};
        by Kosmas, Vergados et al.
\cite{Vergados:1996hs,Kosmas:1997jm,Divari:2000dc};
        the so-called ``{hybrid}'' model of Dimitrov, Engel and Pittel 
\cite{Dimitrov:1995gc}
        and perturbation theory 
	based on calculations of Engel et al.
\cite{Engel:1995gw}.
	For the experimentally interesting nuclear systems 
	$^{29}_{}$Si and $^{73}_{}$Ge 
	very elaborate calculations have been performed  by Ressell et al. 
\cite{Ressell:1993qm}. 
	In the case of $^{73}_{}$Ge 
	a further improved calculation by 
	Dimitrov, Engel and Pittel was carried out 
\cite{Dimitrov:1995gc}
	by suitably mixing variationally determined 
	triaxial Slater determinants. 
	At the present time 
	the necessity for more detailed calculations {\em especially}\/ 
	for the spin-dependent component of the cross sections 
	for heavy nuclei is well motivated.

        To perform modern 
	data analysis in the finite momentum transfer approximation 
	it looks reasonable to use formulae for the  
 	differential event rate 
(Eq.~(\ref{Definitions.diff.rate})) as schematically given below:
\begin{eqnarray}
\label{fit-finite-q}
\label{Definitions.spin.differential.rate}
\frac{dR(\eth,\emx)}{d\ER}
&=& 
{\cal N}(\eth,\emx,\ER,m_\chi)
     \left[ \eta^{}_\SI(\ER,m_\chi) \,\sigma^{p}_\SI
           +\eta^{\prime}_\SD(\ER,m_\chi,\omega) \,  {a_0^2}
     \right ];
\\ \nonumber
{\cal N}(\eth,\emx,\ER,m_\chi) 
&=&  \left[ N_T \frac{c \rho_\chi}{2 m_\chi} 
      \frac{M_A}{\mu_p^2}
      \right]
      \frac{4 \mu_A^2}{\left< q^2_{\max}\right>}
      \langle \frac{v}{c} \rangle  I(\ER)
       \theta (\ER-\eth) \theta(\emx-\ER),
\\ \nonumber
\eta^{}_\SI(\ER,m_\chi)
&=& \left\{ A^2 F^2_\SI(\ER)\right\}
; \\ \nonumber
\eta^{\prime}_\SD(\ER,m_\chi,\omega)
&=& \mu_p^2
\left\{ \frac{4 
}{2J+1} 
	\left(S_{00}(q) + \omega^2\, S_{11}(q) + \omega\, S_{01}(q) 
\right) \right\} .
\end{eqnarray}
       Here the ratio of isovector-to-isoscalar nucleon couplings 
       is $\omega = {a_1}/{a_0}$.
       The detector threshold recoil energy 
       $\eth$ and the maximal available recoil energy $\emx$ 
       ($\eth \le \ER \le \emx$) have been introduced already in Eq. 
(\ref{Definitions.total.rate}).
       In practice, for example with
       an ionization or scintillation signal, one has to 
       take into account the quenching of the recoil energy, when 
       the visible recoil energy is smaller than the real
       recoil energy transmitted by the WIMP to the target nucleus.

       Formulae 
(\ref{fit-finite-q}) allow experimental recoil 
       spectra to be directly described in terms of only {\em three}\ 
\cite{Bednyakov:1994te} 
       (it is rather reasonable to assume 
       $\sigma^{p}_\SI(0)\approx \sigma^{n}_\SI(0)$) 
       independent parameters 
       ($\sigma^{p}_\SI$, $a^2_0$ and $\omega$)
       for any fixed WIMP mass $m_\chi$ and any neutralino composition.
	Comparing this formula with the observed recoil spectra 
	for different targets (Ge, Xe, F, NaI, etc)
	one can directly and simultaneously 
	restrict both isoscalar $c_0$ (via $\sigma^{p}_\SI$) and isovector 
	neutralino-nucleon effective couplings $a_{0,1}$.
	These constraints,
	based on the nuclear spin structure functions for finite $q$, 
	will impose {\em the most 
	model-independent and most accurate restrictions}\ 
	on any SUSY parameter space.
       Contrary to some other possibilities 
       (see, for example, \cite{Bernabei:2003za} and 
\cite{Tovey:2000mm}), 
       this procedure is direct and uses
       as much as possible the results of the 
       accurate nuclear spin structure calculations.

       It is seen from 
Eqs.~(\ref{effectiveSD-cs-pn}) and (\ref{fit-finite-q}) that
        the SD cross sections $\sigma^{p}_{\SD}$ 
	and $\sigma^{n}_{\SD}$ 
	(or equivalently $a^2_{0}$ and $\omega = {a_1}/{a_0}$) are 
        the only two 
	WIMP-nucleon spin variables which can be  
	constrained (or extracted) from DM measurements. 
        Therefore there is no sense to extract from the data 
        (with ``artificial'' twofold ambiguity)
        effective WIMP-nucleon couplings $a^{}_{p}$ and $a^{}_{n}$. 

\section{Cross sections in the effective low-energy MSSM}  
        To estimate the expected direct DM detection rates 
	(with formulae
(\ref{Definitions.diff.rate}),
(\ref{for-toy-mixig})
or (\ref{fit-finite-q}))
        one should calculate cross sections $\sigma^{}_\SI$
        and $\sigma^{}_\SD$ 
	(or WIMP-nucleon couplings $c_0$ and $a^{}_{p,n}$)
	within the framework of 
        some SUSY-based theory or take them 
	from some experimental data (if it is possible).

	To obtain as much as general SUSY predictions
	it appeared  more convenient to work within 
	a phenomenological effective low-energy 
	minimal SUSY model (effMSSM)
	whose parameters are defined 
	directly at the electroweak scale, 
	relaxing completely constraints following from 
	any unification assumption
	(see for example, 
\cite{Mandic:2000jz,Bergstrom:1996cz,Gondolo:2000fh,Bergstrom:2000pn,%
Bottino:2000jx,Ellis:2003ry,Ellis:2003eg,%
Bednyakov:2003wf,Bednyakov:2002dz,Bednyakov:2002js,%
Bednyakov:2002mb,Bednyakov:2000he,Bednyakov:2002ng,%
Bednyakov:1999vh,Bednyakov:1997jr,Bednyakov:1997ax,Bednyakov:1994qa}).
	The effMSSM parameter space is determined by entries of the mass 
	matrices of neutralinos, charginos, Higgs bosons, 
	sleptons and squarks. 
	The list of free parameters includes 
	$\tan\beta$, the ratio
	of neutral Higgs boson vacuum expectation values; 
	$\mu$, the bilinear Higgs parameter of the superpotential;
	$M_{1,2}$, soft gaugino masses; 
	$M_A$, the CP-odd Higgs mass; 
	$m^2_{\widetilde Q}$, $m^2_{\widetilde U}$, $m^2_{\widetilde D}$ 
	($m^2_{\widetilde L}$, $m^2_{\widetilde E}$), 
	squark (slepton) 
	mass parameters squared for the 1st and 2nd generation;        
	$m^2_{\widetilde Q_3}$, $m^2_{\widetilde T}$, $m^2_{\widetilde B}$ 
	($m^2_{\widetilde L_3}$, $m^2_{\widetilde \tau}$), 	
	squark (slepton) mass parameters squared 
	for the 3rd generation; 
	$A_t$, $A_b$, $A_\tau$, soft trilinear 
	couplings for the 3rd generation.
	The third gaugino mass parameter $M_3$ defines the 
	mass of the gluino in the model and is 
	determined by means of the GUT assumption $M_2 = 0.3\, M_3$.
	In the MSSM the lightest neutralino 
	$\chi \equiv \tilde\chi_1^0$
	is a mixture of four superpartners of gauge and Higgs bosons
	(Bino, Wino and two Higgsinos):
$\chi =   N_{11}\widetilde B^0  +N_{12}\widetilde W^0
	+N_{13}\widetilde H_1^0+N_{14}\widetilde H_2^0.
$ 
	Concerning the neutralino composition  
	it is commonly accepted that $\chi$ is mostly gaugino-like if 
	$P\equiv N^2_{11}+N^2_{12}>0.9~$ and  
	Higgsino-like if $P < 0.1$,\ or mixed otherwise.
	
	To constrain the huge effMSSM parameter space and to 
	have reliable predictions for the dark matter experiments
	one usually takes into account available
	information from colliders, astrophysics and rare decays.
	In our previous considerations
\cite{Bednyakov:2003wf,Bednyakov:2002dz,Bednyakov:2002js,%
Bednyakov:2002mb,Bednyakov:2000he,Bednyakov:2002ng,%
Bednyakov:1999vh} 
	the experimental upper limits on sparticle and Higgs masses
	from their non-observations 
\cite{Hagiwara:2002fs,Yao:2006px} were included. 
	Also the limits on the rare $b\rightarrow s \gamma$ decay 
\cite{Alam:1995aw,Abe:2001fi} following 
\cite{Bertolini:1991if,Barbieri:1993av,Buras:1994xp,Ali:1993ct} 
	have been imposed. 

	Furthermore, for each point in the MSSM parameter space (MSSM model) 
	the relic density of the light neutralinos $\Omega_{\chi} h^2$ 
	was evaluated with the code 
\cite{Bednyakov:2002dz,Bednyakov:2002js,Bednyakov:2002ng} based on 
        the code DarkSUSY
\cite{Gondolo:2000ee} with the allowance for 
	all coannihilation channels with 
	two-body final states that can occur between neutralinos, charginos,
	sleptons, stops and sbottoms
	as long as their masses are $m_i<2m_\chi$.
	Two cosmologically interesting regions were considered. 
        One is 
	$0.1< \Omega_\chi h^2  < 0.3$ and the other is 
	the WMAP-inspired region $0.094< \Omega_\chi h^2  < 0.129$ 
\cite{Spergel:2003cb,Bennett:2003bz}.
	The possibility the LSP to be not the only 
	DM candidate, with much smaller relic density 
	$0.002< \Omega h^2  < 0.1$ is also taken into account. 
         Further details can be found in 
\cite{Bednyakov:2004be}.
	In numerical studies of 
\cite{Bednyakov:2000he,Bednyakov:2000uw,Bednyakov:2002dz,%
Bednyakov:2002js,Bednyakov:2002ng}
	the parameters of the effMSSM are 
	randomly varied in the following intervals: 
\begin{eqnarray} \nonumber
&-1{\rm ~TeV} < M_1 < 1{\rm ~TeV}, \quad
-2{\rm ~TeV} < M_2, \mu, A_t < 2{\rm ~TeV},&
\\ \label{Scan}
&1 < \tan\beta < 50, \quad 60{\rm ~GeV} < M_A < 1000{\rm ~GeV},&\\
\nonumber
&10{\rm ~GeV}^2 < m^2_{Q_{}},
m^2_{L}, m^2_{Q_3}, m^2_{L_3}<10^6{\rm ~GeV}^2.&
\end{eqnarray}
	For the other sfermion mass parameters as before in 
\cite{Bednyakov:2003wf,Bednyakov:2002dz,Bednyakov:2002js,%
Bednyakov:2002mb,Bednyakov:2000he,Bednyakov:2002ng,Bednyakov:1999vh} 
	we used the relations 
	$m^2_{\widetilde U_{}} = m^2_{\widetilde D_{}} 
	= m^2_{\widetilde Q_{}}$, 
	$m^2_{\widetilde E_{}} = m^2_{\widetilde {L}}$, 
	$m^2_{\widetilde T} = m^2_{\widetilde B} = m^2_{\widetilde{Q}_3}$,  
	$m^2_{\widetilde{E}_{3}} = m^2_{\widetilde{L}_3}$.
	The parameters $A_b$ and $A_\tau$ are fixed to be zero.
	We consider the domain of the MSSM parameter space, 
	in which we perform our scans, as quite spread and natural. 

\begin{figure}[!ht] 
\begin{picture}(100,115)
\put(-13,-37){\includegraphics{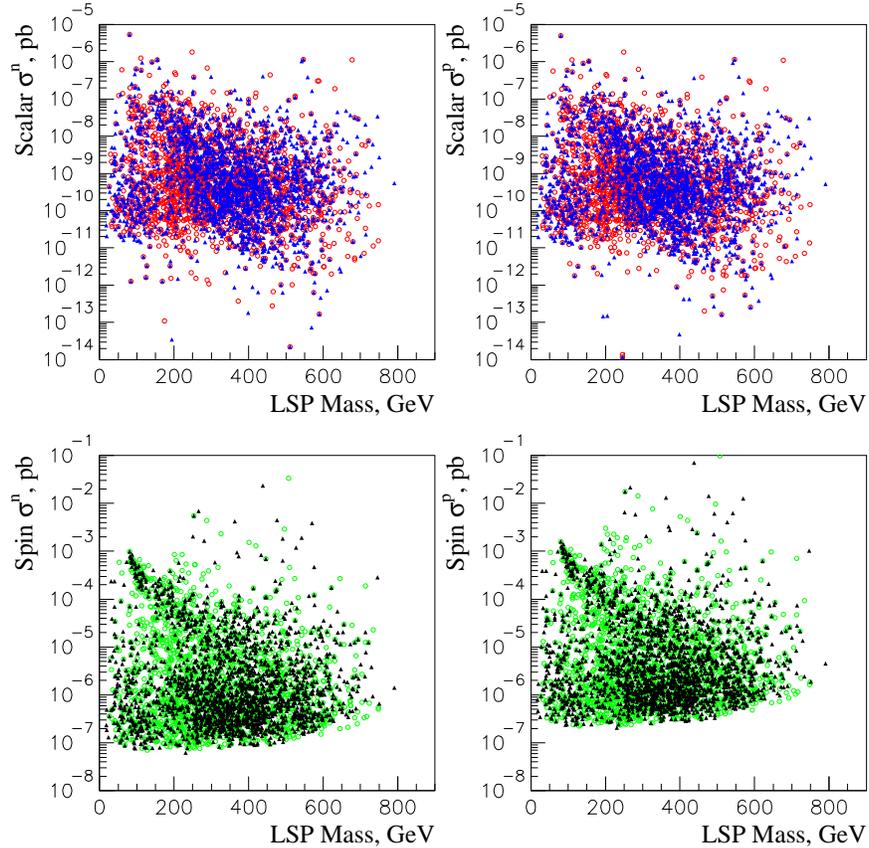}}
\end{picture}
\caption{Cross sections of spin-dependent and spin-independent
	interactions of WIMPs with proton and neutron.
	Filled triangles (open circles)
	correspond to relic neutralino density 
	$0.1 < \Omega_\chi h^2<0.3$ 
	($0.025 < \Omega_\chi h^2<1$).
\label{CrossSections}
From \cite{Bednyakov:2000he,Bednyakov:2000uw,Bednyakov:2002dz,%
Bednyakov:2002js,Bednyakov:2002ng}. 
}
\end{figure}

	Typical WIMP-nucleon cross sections 
	of both spin (SD) and scalar (SI)
	interactions as function of the WIMP mass 
	are shown as scatter plots in  
Fig.~\ref{CrossSections}. 
	In the figure open circles correspond to cross sections
	calculated under the old assumption that  
	$0.025 < \Omega_\chi h^2<1$.
	Filled triangles give the same cross section but the constraint 
	on the flat and accelerating universe is imposed by 
	$0.1 < \Omega_\chi h^2<0.3$.
	One can see that the reduction of the allowed domain
	for the relic density does not significantly affect
	spin-dependent and the spin-independent WIMP-nucleon cross sections.
	The different behavior of SD and SI cross sections with 
	mass of the LSP can be seen from the plots.
	There is a more stringent 
	lower bound for the spin-dependent cross section.
	It is at a level of $10^{-7}\,$~pb.

\begin{figure}[!ht] 
\begin{picture}(100,120)
\put(-20,-5){\includegraphics{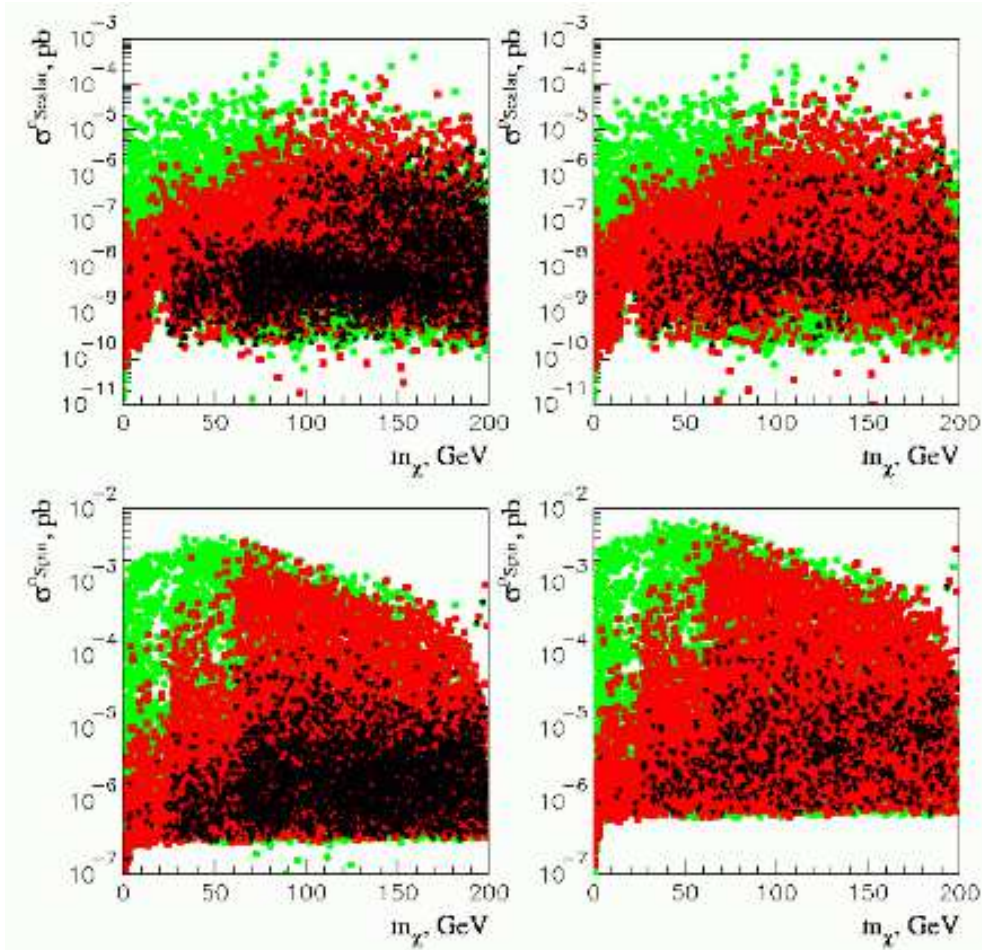}}
\end{picture}
\caption{Cross sections of the spin-dependent (spin) and 
	the spin-independent (scalar)
	interactions of WIMPs with the proton and the neutron.
	Filled green circles correspond to the relic neutralino density 
	$0< \Omega_\chi h^2_0<1$,
	red squares correspond to the sub-dominant relic neutralino
	contribution $0.002 < \Omega_\chi h^2_0<0.1$ 	
	and black triangles correspond to the relic neutralino density 
	$0.1 < \Omega_\chi h^2_0<0.3$ (left panel) and to 
	the WMAP relic density
	$0.094 < \Omega_\chi h^2_0<0.129$ (right panels). 
\label{CrossSections-vs-lsp}}
\end{figure}

        For more accurate investigation of the DAMA-inspired 
	domain of the lower masses of the LSP ($m_\chi < 200$~GeV) in 
\cite{Bednyakov:2004be}
        both $\sigma^{}_\SD$ and $\sigma^{}_\SI$ have also been
	calculated within the effMSSM.
	To this end 
	the intervals of the randomly scanned MSSM parameter space in 
\cite{Bednyakov:2004be} were narrowed: 
\begin{eqnarray}\label{Scanning}
-200{\rm ~GeV} < M_1 < 200{\rm ~GeV}, \ \
-1{\rm ~TeV} < M_2, \mu < 1{\rm ~TeV}, \ \
50{\rm ~GeV} < M_A < 500{\rm ~GeV}. 
\end{eqnarray}
        The results of these evaluations are shown as scatter plots 
in Fig.~\ref{CrossSections-vs-lsp}, which 
            is the WIMP low-mass update of 
Fig.~\ref{CrossSections}. 
	In 
Fig.~\ref{CrossSections-vs-lsp}
        filled green circles correspond to cross sections
	calculated when the neutralino relic density 
	should just not overclose the Universe 
	($0.0<\Omega_\chi h^2_0<1.0$).
	Filled red squares show the same cross sections 
	when one assumes the relic neutralinos to be not the only 
	DM particles and give only a sub-dominant 
	contribution to the relic density  
 	$0.002 < \Omega_\chi h^2_0<0.1$.
 	In the left panel of 
Fig.~\ref{CrossSections-vs-lsp}
	these cross sections are shown with the  black triangles
	corresponding to the case when the relic neutralino density 
	is in the bounds previously associated with a so-called 
	flat and accelerating Universe
	$0.1 < \Omega_\chi h^2_0<0.3$.
	The black triangles in 
Fig.~\ref{CrossSections-vs-lsp} (right panel) 
	correspond to the the WMAP and SDSS
\cite{Spergel:2003cb,Bennett:2003bz}
	constraint on the matter relic density   
	$0.094 < \Omega_\chi h^2_0<0.129$ imposed in 2004. 
	Despite a visible reduction   
	of the allowed domain for the relic density due to the 
	WMAP+SDSS result the upper bounds for the 
	spin-dependent and the spin-independent WIMP-nucleon cross section 
	are not significantly affected.

	Finally it is perhaps the right place here, 
	to comment the following.
	Unfortunately the MSSM parameter space is huge and to obtain 
	some reliable feeling, concerning, for example, 
	the expected rate of dark matter detection
	when all relevant experimental and cosmological 
	constraints are taken into account, one 
	has nothing but this statistical numerical method
(see for example, 
\cite{Drees:1993bu,Jungman:1996df,Bednyakov:1999yr,%
Bednyakov:1996yt,Bednyakov:1997ax,Bednyakov:1997jr,%
Bottino:2000jx,Bottino:1997eu}).
 	This method allows lower and upper bounds for
	any observable to be estimated, 
	and to  make conclusions about the prospects 
	for dark matter detection with present 
	or future high-accuracy dark matter detectors. 
	The larger the amount of points which confirm 
	such a conclusion the better. 
	The conclusions we made here are based on hundreds of thousands of
	points which passed all constraints.
	Of course, we have no proved protection against
	peculiar choices of parameters which
	could lead to some cancellation and to 
	small cross sections even if Higgs masses are small. 
	Nevertheless,
	the probability of these choices is very small
	(about 1/100000), otherwise we should already meet them
	with our random scanning. 
	On the  other side, 
	if these peculiar choices exist and one day would manifest 
	themselves, 
	this would be a very interesting puzzle, 
	because it would be some kind of fine tuning of parameters, 
	which requires strong further development of
	our understanding of the theory
\cite{Bednyakov:1999vh}.       

\section{One-coupling dominance approach} 
\begin{figure}[!ht] 
\begin{picture}(100,120) 
\put(-2,-5){\includegraphics{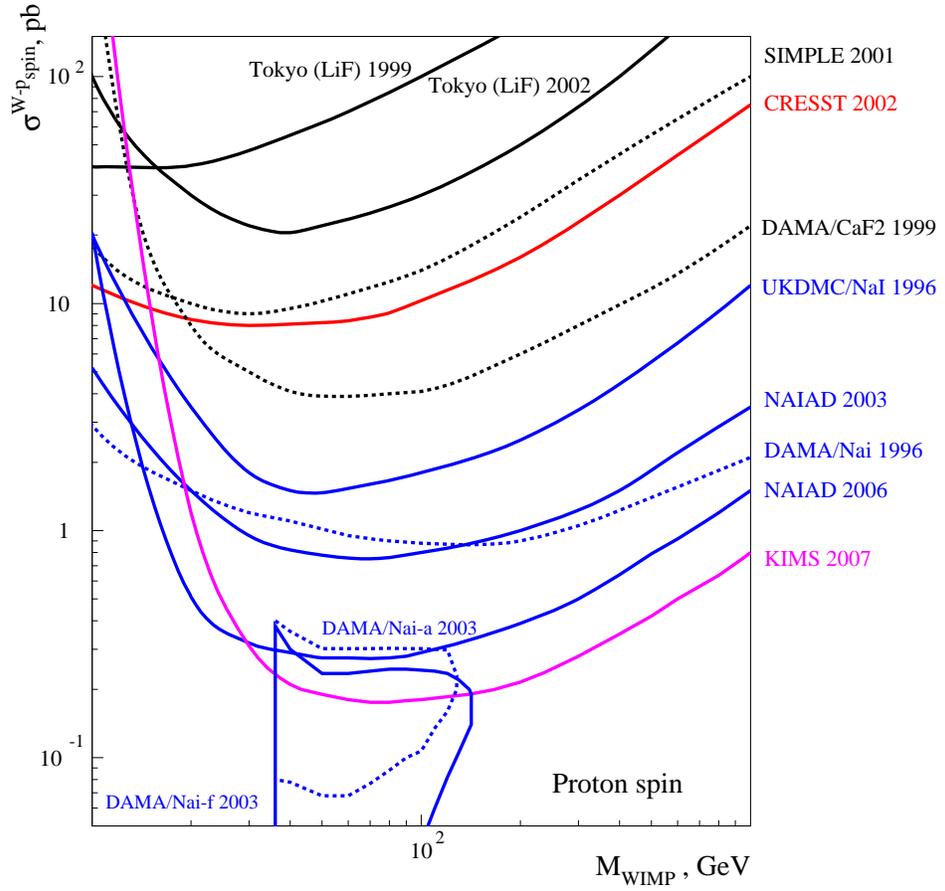}}
\end{picture}
\caption{Exclusion curves for the spin-dependent WIMP-proton cross sections
        $\sigma^{p}_{{\rm SD}}$ as a function of the WIMP mass.
        All curves, except the NAIAD and Tokio-LiF, are 
        obtained in the one-coupling dominance approach 
        with $\sigma^{}_{\rm SI}=0$ and $\sigma^{n}_{{\rm SD}}=0$.
        The DAMA/NaI-a(f) contours for the WIMP-proton 
	SD interaction in $^{127}$I
        are obtained on the basis of the positive 
        signature of annual modulation within the 
        framework of the mixed scalar-spin coupling approach
\cite{Bernabei:2003za,Bernabei:2001ve}. 
For details see \cite{Bednyakov:2004be}.}
\label{Spin-p} 
\end{figure}

\begin{figure}[!ht] 
\begin{picture}(100,100)
\put(-8,-3){\includegraphics{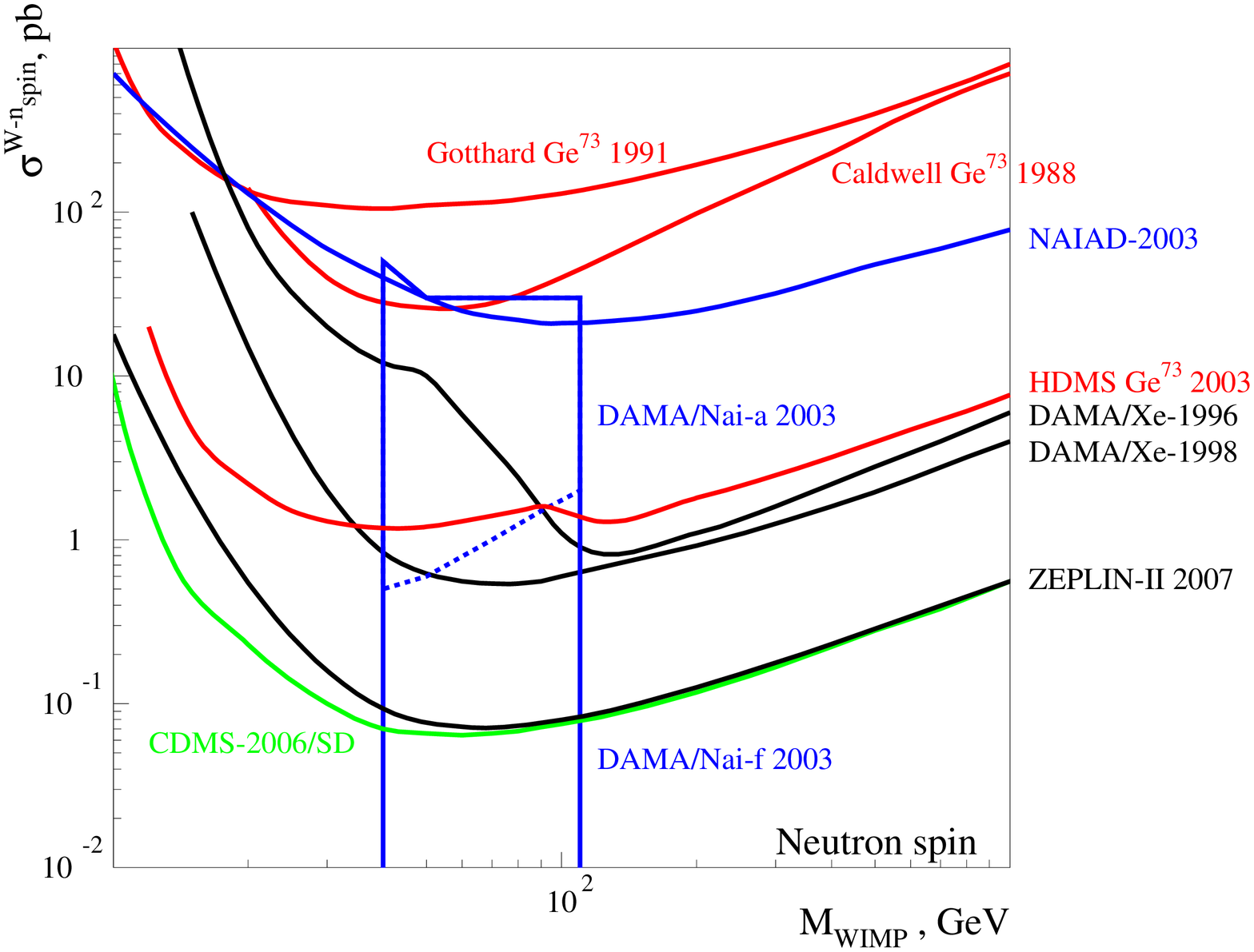}}
\end{picture}
\caption{Exclusion curves for the spin-dependent WIMP-neutron cross sections
        ($\sigma^{n}_{{\rm SD}}$ versus the WIMP mass).
        The DAMA/NaI-a(f) contours for the WIMP-neutron SD interaction
        (subdominating in $^{127}$I) 
        are obtained from the relevant figures of   
\cite{Bernabei:2003za,Bernabei:2001ve}. 
        Note that the NAIAD curve 
\cite{Ahmed:2003su} here corresponds to the 
        WIMP-neutron SD interaction subdominant for $^{127}$I. 
        The WIMP-proton SD interaction dominates for this nucleus.
        The curve was obtained in the approach of 
\cite{Tovey:2000mm}. 
        It is much weaker in comparison with 
	both the DAMA/Xe and HDMS-2003 curves. 
(For more details see \cite{Bednyakov:2004be}
and Fig.~\ref{mixed-couplings}).}
\label{Spin-n} 
\end{figure}

        From the definitions of SD and SI WIMP-nucleus and
	WIMP-nucleon cross sections  
(Eqs.~(\ref{Definitions.scalar.zero.momentum})--%
(\ref{effectiveSD-cs-pn}),
(\ref{SI-cs-Deff.at-zero}) and
(\ref{SD-cs-Deff.at-zero}))
        one can conclude  that          
        the spin observables in DM search 
	give us two independent constraints on a SUSY model via
	$\sigma^{p}_{{\rm SD}}(0)$ and $\sigma^{n}_{{\rm SD}}(0)$, 
	or, equivalently, via ${a}^{}_{p}$  and ${a}^{}_{n}$
	(or  ${a}^{}_{0}$  and ${a}^{}_{1}$).
        These constraints are usually presented in the form
	of exclusion curves obtained with different target nuclei 
	and recalculated in terms of nuclear-independent 
	$\sigma^{p}_{{\rm SD}}(0)$ 
(see for example, Fig.~\ref{Spin-p})
	and $\sigma^{n}_{{\rm SD}}(0)$ 
(see for example, Fig.\ref{Spin-n}).
       For a fixed mass of the WIMP the cross sections
       of SI or SD elastic WIMP-nucleon interaction
       located above these curves are excluded.  

\begin{figure}[!ht] 
\begin{picture}(60,102) 
\put(-34,-3){\includegraphics{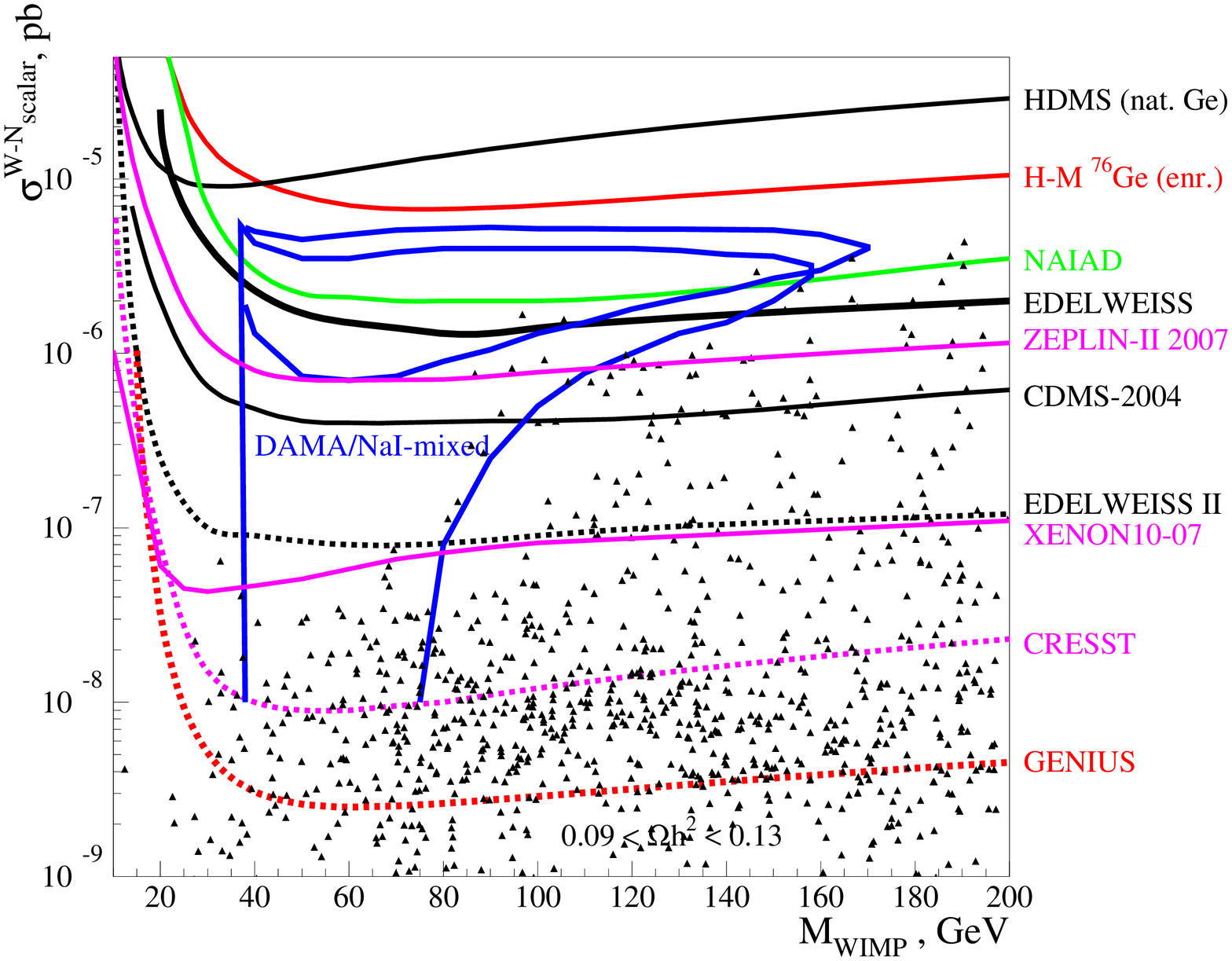}}
\end{picture}
\caption{WIMP-nucleon cross section ($\sigma^{p}_{{\rm SI}}(0)$) 
        limits in pb for spin-independent (scalar) interactions as 
        a function of the WIMP mass in GeV. 
        Shown are contour lines for some of the present experimental limits 
        (solid lines) and some of projected experiments (dashed lines). 
        All curves are obtained in 
        the one-coupling dominance approach with $\sigma^{}_{\rm SD}=0$.
        For example, 
        the closed DAMA/NaI contour corresponds to complete neglection 
        of SD WIMP-nucleon interaction.
        The open DAMA contour is obtained in 
\cite{Bernabei:2003za} 
        with the assumption that $\sigma^{}_{\rm SD}=0.08$ pb $> 0$.
	Theoretical expectations for 
	$\sigma^p_{\rm SI}$ in the effMSSM from 
\cite{Bednyakov:2004be}
	are also shown by scatter plots for a relic neutralino density
	$0.09 < \Omega_\chi h^2_0<0.13$ (black triangles)
(see \cite{Bednyakov:2004be}).        }
\label{Scalar-2004}
\label{Scalar-2003}  
\end{figure}

	This simple presentation 
	 allows one to compare directly 
	 sensitivities of DM experiments with different nuclear targets.
	 At the current level of accuracy (when 
	 $f^{(p)}_q \approx f^{(n)}_q$ and 
	 $\sigma^{p}_{\SI}(0) \approx \sigma^{n}_{\SI}(0)$, see 
Fig.~\ref{CrossSections}) 
	 there is {\em only one}\/ 
	 constraint for a WIMP-nucleon cross section 
(see Fig.~\ref{Scalar-2003}) 
	from spin-independent DM search experiments. 
        Indeed, 
        for the spin-zero nuclear target the experimentally 
	measured event rate 
(Eq.~(\ref{Definitions.diff.rate})) 
	of direct DM particle detection, via formula 
(\ref{Definitions.cross.section}) 
	is connected with the zero-momentum WIMP-proton(neutron)
        cross section 
(\ref{Definitions.scalar.zero.momentum}). 
        The zero momentum scalar WIMP-proton(neutron) cross section 
        $\sigma^{p}_{{\rm SI}}(0)$ can be expressed through 
        effective neutralino-quark couplings ${\cal C}_{q}$
(\ref{Definitions.effective.lagrangian}) by means of expression 
(\ref{Definitions.scalar.zero.cs}).
        These couplings ${\cal C}_{q}$ (as well as ${\cal A}_{q}$) 
        can be directly connected with the
        fundamental parameters of a SUSY model 
        such as $\tan \beta$, $M_{1,2}$, $\mu$, masses 
        of sfermions and Higgs bosons, etc. 
        Therefore experimental limitations on the SI 
	neutralino-nucleon cross section
        supply us with a constraint on the fundamental parameters
        of an underlying SUSY model.
        In the case of the SD 
	WIMP-nucleus interaction
        from a measured differential rate 
Eq.~(\ref{Definitions.diff.rate}) one first extracts 
        a limitation for $\sigma^{A}_{{\rm SD}}(0)$, 
        and therefore has in principle two constraints
\cite{Bednyakov:1994te}
        for the neutralino-proton $a^{}_p$ and  
        neutralino-neutron $a^{}_n$ effective spin couplings, 
	as follows from relation 
(\ref{Definitions.spin.zero.momentum}).
        From
Eq.~(\ref{Definitions.spin.zero.momentum})
        one can also see that contrary to the SI 	case
(\ref{Definitions.scalar.zero.momentum}) there is, in general, 
	no factorization of the nuclear structure
        for $\sigma^A_{\rm SD}(0)$.
        Both proton $\langle{\bf S}^A_{p}\rangle$
        and neutron $\langle{\bf S}^A_{n}\rangle$
        spin contributions simultaneously enter into formula 
(\ref{Definitions.spin.zero.momentum})
        for the SD WIMP-nucleus cross section $\sigma^A_{\rm SD}(0)$.
        Nevertheless, for 
        the most interesting isotopes either $\langle{\bf S}^A_{p}\rangle$ 
        or $\langle{\bf S}^A_{n}\rangle$ dominates
        ($\langle{\bf S}^A_{n(p)}\rangle \ll \langle{\bf S}^A_{p(n)}\rangle$)
\cite{Bednyakov:2004be,Bednyakov:2004xq}.

        In earlier considerations 
\cite{Smith:1990kw,Engel:1992bf,Ellis:1988sh,%
Goodman:1985dc,Drukier:1986tm,Engel:1989ix}
        one reasonably assumed that the nuclear spin was carried by the ``odd''
        unpaired group of protons or neutrons and only one of either 
        $\langle{\bf S}^A_n\rangle$ or $\langle{\bf S}^A_p\rangle$ was 
        non-zero.
        In this case all possible non-zero-spin target nuclei can 
        be classified into n-odd and p-odd groups.
        Following this classification,
        the current experimental situation
        for the spin-dependent WIMP-{\bf proton} 
        and WIMP-{\bf neutron} cross sections is 
        naturally presented separately in 
Fig.~\ref{Spin-p} and
Fig.~\ref{Spin-n}.
        The DAMA/NaI-a(f) contours for the WIMP-proton SD interaction 
        (dominating in $^{127}$I) obtained on the basis of the positive 
        signature of the annual modulation (closed contour)
\cite{Bernabei:2003za} and within the mixed coupling framework (open contour)
\cite{Bernabei:2001ve} are also presented in 
Fig.~\ref{Spin-p}. 
        Similarly, the DAMA/NaI-a(f) 
\cite{Bernabei:2003za} contours for the WIMP-neutron SD interaction
        (subdominant in $^{127}$I) are given in
Fig.~\ref{Spin-n}. 
        There are also 
        exclusion curves for the SD cross section from the CDMS 
\cite{Akerib:2004fq} and EDELWEISS 
\cite{Sanglard:2004kb} experiments
        with natural-germanium bolometric detectors
        (due to the small Ge-73 admixture). 

        To compare experimental data with theoretical estimations
	in the effMSSM 
\cite{Bednyakov:2004be} 
	one can superimpose the scatter plots
        for the SD and SI LSP-proton and LSP-neutron cross sections 
        (from  
Fig.~\ref{CrossSections-vs-lsp} or
Fig.~\ref{CrossSections}) 
in Figs.~\ref{Spin-p}, \ref{Spin-n}, and \ref{Scalar-2004}.
           This is a traditional way to perform this comparison.
	One can easily see that both calclulated 
        SD LSP-proton and LSP-neutron 
        cross sections fall below the frames of 
Figs.~\ref{Spin-p} and \ref{Spin-n}, respectively.
        In particular, this means that 
	experimental data 
        (avaliable at present in the form of exclusion curves)  
	do not allow one to restrict the SUSY LSP-nucleon
	spin couplings.
	This is not the case for the SI WIMP-nucleon coupling.
	The scattered points (black triangles) for 
	$\sigma^p_{\rm SI}$ calculated in the effMSSM are clear seen 
in Fig.~\ref{Scalar-2004}.
         Some of these points are already excluded by the 
	 DM measuremnts.
       
	Nevertheless,
        one would like to note that, for example, the calculated scatter 
        plots for $\sigma^p_{\rm SD}$ 
(Fig.~\ref{Spin-p})
        are obtained without any assumption
        of $\sigma^n_{\rm SD}=0$ (and $\sigma^p_{\rm SI}=0$), 
        but the experimental exclusion curves for $\sigma^p_{\rm SD}$ 
        were traditionally extracted from the data 
        with the spin-neutron (and scalar) contribution
        fully neglected,
        i.e. under the assumption that 
        $\sigma^n_{\rm SD}=0$ (and $\sigma^p_{\rm SI}=0$).
        This 
{\bf one-spin-coupling dominance scheme} 
        (always used before the new approaches were proposed in
\cite{Tovey:2000mm} and in 
\cite{Bernabei:2003za,Bernabei:2000qi,Bernabei:2003wy}) 
        gave a bit too pessimistic exclusion curves,
        but allowed direct and {\em correct}\ 
	comparison of sensitivities 
        for different DM search experiments.
        More stringent constraints on $\sigma^p_{\rm SD}$ can be obtained 
\cite{Tovey:2000mm,Bernabei:2003za,Bernabei:2000qi,Bernabei:2003wy}
        by assuming both 
        $\sigma^p_{\rm SD}\neq 0$ and $\sigma^n_{\rm SD}\neq 0$
        (although the contribution of the neutron spin is
        usually very small because
        $\langle{\bf S}^A_{n}\rangle \ll \langle{\bf S}^A_{p}\rangle$).
        Therefore a direct 
        comparison of an old-fashioned exclusion curve 
        with a new one could in principle mislead one to a wrong
	conclusion about better sensitivity of the new experiment. 

        The same conclusion on the one-coupling dominance approach  
        to a great extent concerns 
\cite{Bernabei:2003za,Bernabei:2003wy,Bednyakov:2004be,Bednyakov:2005qp}
        the direct comparison of the {\em old}\/ SI exclusion curves 
        (obtained with $\sigma^{}_{\rm SD}=0$)
        with the {\em new}\/ SI exclusion curves 
        (obtained with         $\sigma^{}_{\rm SD}>0$)
        as well as with the results of the SUSY calculations.
        One can see from 
Fig.~\ref{Scalar-2004} that the {\em new-type}\/ DAMA/NaI open contour
        (when $\sigma^{}_{\rm SD}>0$)
        is in agreement with the best exclusion curves of the  
        CDMS and EDELWEISS as well as with SUSY calculations.
        One knows that both of these latter experiments have natural
        germanium (almost purely spinless) 
        as a target and therefore have only little sensitivity to the 
        spin-dependent WIMP-nucleon couplings 
        (for them $\sigma^{}_{\rm SD}\simeq 0$). 
        Therefore, these experiments exclude only the pure 
        SI interpretation of the DAMA annual modulation signal 
\cite{Akerib:2004fq,Chardin:2004ry,Kurylov:2003ra,Copi:2000tv,Copi:2002hm}.
        The statement that this DAMA result 
        is {\em completely}\ excluded by the results of these 
        cryogenic experiments and 
        is inconsistent with the SUSY interpretation (see, for example,
\cite{Drees:2004db}) 
        is simply wrong (see also discussions in 
\cite{Kurylov:2003ra,Gelmini:2004gm,Gondolo:2005hh,Gelmini:2005fb}).

        The event-by-event CDMS and EDELWEISS background discrimination
        (via simultaneous charge and phonon signal measurements) is 
        certainly very important. 
        Nevertheless the DAMA annual signal modulation is one of 
	a few available {\em positive}\/ signatures of 
        WIMP-nucleus interactions and the importance 
        of its observation goes far beyond a ``simple'' 
        background reduction. 
        Therefore, to completely exclude the DAMA result, a new experiment, 
        being sensitive to the modulation signal, 
        would have to confirm or exclude this 
        modulation signal on the basis of 
        the same or much better statistics. 

         Furthermore, taking seriously 
         the positive DAMA result together with 
         the negative results of the CDMS and EDELWEISS
         as well as the results of
\cite{Savage:2004fn}
         one can arrive at a conclusion about simultaneous
         existence and 
         importance of both SD and SI WIMP-nucleus interactions.

\enlargethispage{-\baselineskip}

\section{Mixed spin-scalar WIMP-nucleon interactions}
        The accurate calculations of spin nuclear structure 
\cite{Iachello:1991ut,Engel:1991wq,Pacheco:1989jz,%
      Engel:1992qb,Engel:1995gw,Dimitrov:1995gc,%
      Ressell:1993qm,Ressell:1997kx,%
      Vergados:1996hs,Kosmas:1997jm,Divari:2000dc,Bednyakov:2004xq}
        demonstrate that contrary to the simplified odd-group approach both
        $\langle{\bf S}^A_{p}\rangle$ and $\langle{\bf S}^A_{n}\rangle$ 
        differ from zero, but nevertheless 
        one of these spin quantities always dominates
        ($\langle{\bf S}^A_{p}\rangle \ll \langle{\bf S}^A_{n}\rangle$, or
         $\langle{\bf S}^A_{n}\rangle \ll \langle{\bf S}^A_{p}\rangle$). 
	It follows form 
Eq.~(\ref{Definitions.spin.zero.momentum.Tovei}), that 
{\it if together}\ with the dominance like 
        $|\langle{\bf S}^A_{p(n)}\rangle| \ll |\langle{\bf S}^A_{n(p)}\rangle|$
        one would have the WIMP-proton and WIMP-neutron couplings
        of the same order of magnitude
        ({\it not} $|a_{n(p)}|\! \ll\! |a_{p(n)}|$), 
        the situation could look like that in the odd-group model
	and one could safely (at the current level of accuracy)
	neglect a sub-dominant spin contribution in the data analysis
	due to the inequality:
$|a_p\langle {\bf S}^A_p\rangle| \ll |a_n\langle {\bf S}^A_n\rangle|$.
        Nevertheless it was shown in 
\cite{Tovey:2000mm}
        that in the general SUSY model one can meet right 
	a case when 
        $a_{n(p)}\! \ll\! a_{p(n)}$
	and the proton and neutron spin contributions are strongly mixed, 
	i.e. 
	$|a_p\langle {\bf S}^A_p\rangle| \approx 
	|a_n\langle {\bf S}^A_n\rangle|$.

        To separately constrain the SD proton and neutron contributions 
	at least two new approaches appeared
	in the literature
\cite{Tovey:2000mm,Bernabei:2001ve}.
        As the authors of 
\cite{Tovey:2000mm} claimed, their method 
        has the advantage that the limits on individual 
        WIMP-proton and WIMP-neutron SD cross sections 
        for a given WIMP mass can be combined 
        to give a model-independent limit 
        on the properties of WIMP scattering 
        from both protons and neutrons in the target nucleus. 
        The method relies on the assumption that the 
        WIMP-nuclear SD cross section can be presented in the form
$\displaystyle 
        \sigma^A_{\rm SD}(0) \!=\!
        \left( \sqrt{\sigma^p_{\rm SD}|^{}_A} \pm 
       \sqrt{\sigma^n_{\rm SD}|^{}_A} \right)^2$,
        where $\sigma^p_{\rm SD}|^{}_A$
          and $\sigma^n_{\rm SD}|^{}_A$ are auxiliary 
        quantities, not directly connected with measurements.
        Furthermore, to extract a constraint on
        the {\em sub-dominant}\ WIMP-proton spin 
        contribution one should assume 
	the proton contribution dominance 
        for a nucleus whose spin is almost completely 
        determined by the neutrons. 
	From one side,	this may look almost useless, 
        especially because these sub-dominant 
        constraints are always much weaker 
        than the relevant constraints 
        obtained directly with a proton-odd group target
	(one can compare, for example, the
	restrictive potential of the NAIAD exclusion curves in 
Figs.~\ref{Spin-p} and \ref{Spin-n}). 
	From another side, the 
        very large and very small ratios $\sigma_p/\sigma_n
	\sim |a_{p}|/|a_{n}|$ obtained in 
\cite{Tovey:2000mm} correspond to neutralinos which are
	extremely pure gauginos. 
	In this case $Z$-boson exchange in SD interactions is absent 
	and only sfermions give contributions to the SD cross sections. 
	This is a very particular (fine-tuning) case
	which is hardly to be 
	in agreement with the present SUSY search experiments.
	Following an analogy between neutrinos and neutralinos
	one could assume that neutralino couplings
	with the neutron and the proton should not be very different
	and one could expect preferably 
        $|a_n|/|a_p| \approx O(1)$.
	The relation  
        $|a_n|/|a_p| \approx O(1)$ 
        was checked in
\cite{Bednyakov:2002dz,Bednyakov:2003wf} for large LSP masses. 
        For relatively low LSP masses $m_\chi < 200$~GeV in effMSSM 
\cite{Bednyakov:1999vh,Mandic:2000jz,Bergstrom:1996cz,Gondolo:2000fh,Bednyakov:1998is,Bergstrom:2000pn,Bottino:2000jx}
        the $a_n$-to-$a_p$ ratio is located within the bounds
\cite{Bednyakov:2004be}:
\begin{equation}
\label{an2ap-ratio-bounds}
0.5 < \left|\frac{a_n}{a_p} \right| <  0.8.
\end{equation}
        Therefore in the model the couplings are almost the same
        and one can safely neglect
	the $\langle{\bf S}^A_{p(n)}\rangle$-spin 
        contribution in the analysis of the 
	DM data for a nuclear target with 
	$\langle{\bf S}^A_{p(n)}\rangle \ll 
         \langle{\bf S}^A_{n(p)}\rangle$.

	Furthermore, when one compares in the same figure
	an exclusion curve for SD WIMP-proton coupling
	obtained without sub-dominant SD WIMP-neutron contribution 
	and without SI contribution (all curves in 
Fig.~\ref{Spin-p} except the one for NAIAD
\cite{Ahmed:2003su} and the one for Tokyo-LiF
\cite{Miuchi:2002zp}), 
	with a curve from the approach of 
\cite{Tovey:2000mm}, when the sub-dominant contribution is included
	(the NAIAD and Tokyo-LiF curves in 
Fig.~\ref{Spin-p}),
	one {\it ``artificially''}\ improves the sensitivity 
	of the {\it latter}\ curves 
	(NAIAD or Tokyo-LiF) in comparison with the former ones.
	To be consistent and for reliable comparison
	of sensitivities of these experiments, 
	one should, at least, coherently recalculate
	all previous curves in the new manner. 
	This message was clearly stressed in 
\cite{Bernabei:2003za}. 

	The same arguments are true for the results 
	of the SIMPLE experiment 
\cite{Giuliani:2003nf} and search for DM with NaF bolometers   
\cite{Takeda:2003km} where the SI contribution seems also 
        to be completely ignored.
	Although $^{19}$F has the best properties 
	for investigation of WIMP-nucleon spin-dependent interactions  
(see, for example
\cite{Divari:2000dc})
	it is not obvious that one should completely ignore
	spin-independent WIMP coupling with the fluorine. 
	For example, in the relation  
$\sigma^A \sim \sigma^{A,p}_{\rm SD}
         \left[\frac{\sigma^A_{\rm SI}}{\sigma^{A,p}_{\rm SD}}
        +\left(1 + 
	\sqrt{\frac{\sigma^{A,n}_{\rm SD}}{\sigma^{A,p}_{\rm SD}}}
        \right)^2 
        \right]
$	which follows from 
Eqs.~(\ref{Definitions.scalar.zero.momentum})--%
(\ref{Definitions.spin.zero.momentum.Bernabei}),
	it is not a priori clear that 
        $\frac{\sigma^A_{\rm SI}}{\sigma^{A,p}_{\rm SD}} \ll
	\frac{\sigma^{A,n}_{\rm SD}}{\sigma^{A,p}_{\rm SD}}
	$,
        i.e. the SI WIMP-nucleus interaction is much weaker than
        the sub-dominant SD WIMP-nucleus one.
	At least for isotopes with atomic number $A>50$
\cite{Bednyakov:1994qa,Jungman:1996df}
	the neglection of the SI contribution would be a larger 
        mistake than the neglection of the  
        sub-dominant SD WIMP-neutron contribution,  
	when the SD WIMP-proton interaction dominates.

        Therefore we would like to note that 
        the ``old'' odd-group-based approach to 
        analysis of the SD data from experiments with heavy enough
        targets (for example, Ge-73) is still quite suitable, 
        especially when it is not obvious that 
        (both) spin couplings dominate over the scalar one.

        From measurements with $^{73}$Ge one can extract, in principle, 
        not only the dominant constraint for WIMP-nucleon coupling
        $a_n$ (or $\sigma_{\rm SD}^{n}$) 
        but also the constraint for the sub-dominant WIMP-proton coupling
        $a_p$ (or $\sigma_{\rm SD}^{p}$) using the approach of 
\cite{Tovey:2000mm}.
        Nevertheless, the latter constraint will be much weaker
        in comparison with the constraints from p-odd group
        nuclear targets, like $^{19}$F or I.  
        This fact is illustrated by the NAIAD (NaI, 2003) curve in 
Fig.~\ref{Spin-n}, which corresponds to the sub-dominant
        WIMP-neutron spin contribution 
        extracted from the p-odd nucleus $^{127}$I. 

         Another approach for the mixed spin-scalar coupling 
	 data presentation, of Bernabei et al.
\cite{Bernabei:2001ve}, 
	is based on an introduction of the so-called effective 
	SD nucleon cross section $\sigma^{pn}_{\rm SD}(0)$
	($\sigma_{{\rm SD}}$ in 
\cite{Bernabei:2003za,Bernabei:2001ve}) 
	and coupling mixing angle $\theta$
(see Eq.~(\ref{effectiveSD-cs}))
	instead of $\sigma^{p}_{\rm SD}(0)$ and
	$\sigma^{n}_{\rm SD}(0)$.
        With these definitions the SD WIMP-proton and
        WIMP-neutron cross sections are given by relations
(\ref{effectiveSD-cs-pn}).

\begin{figure}[!ht] 
\begin{picture}(100,80)
\put(-37,0){\includegraphics{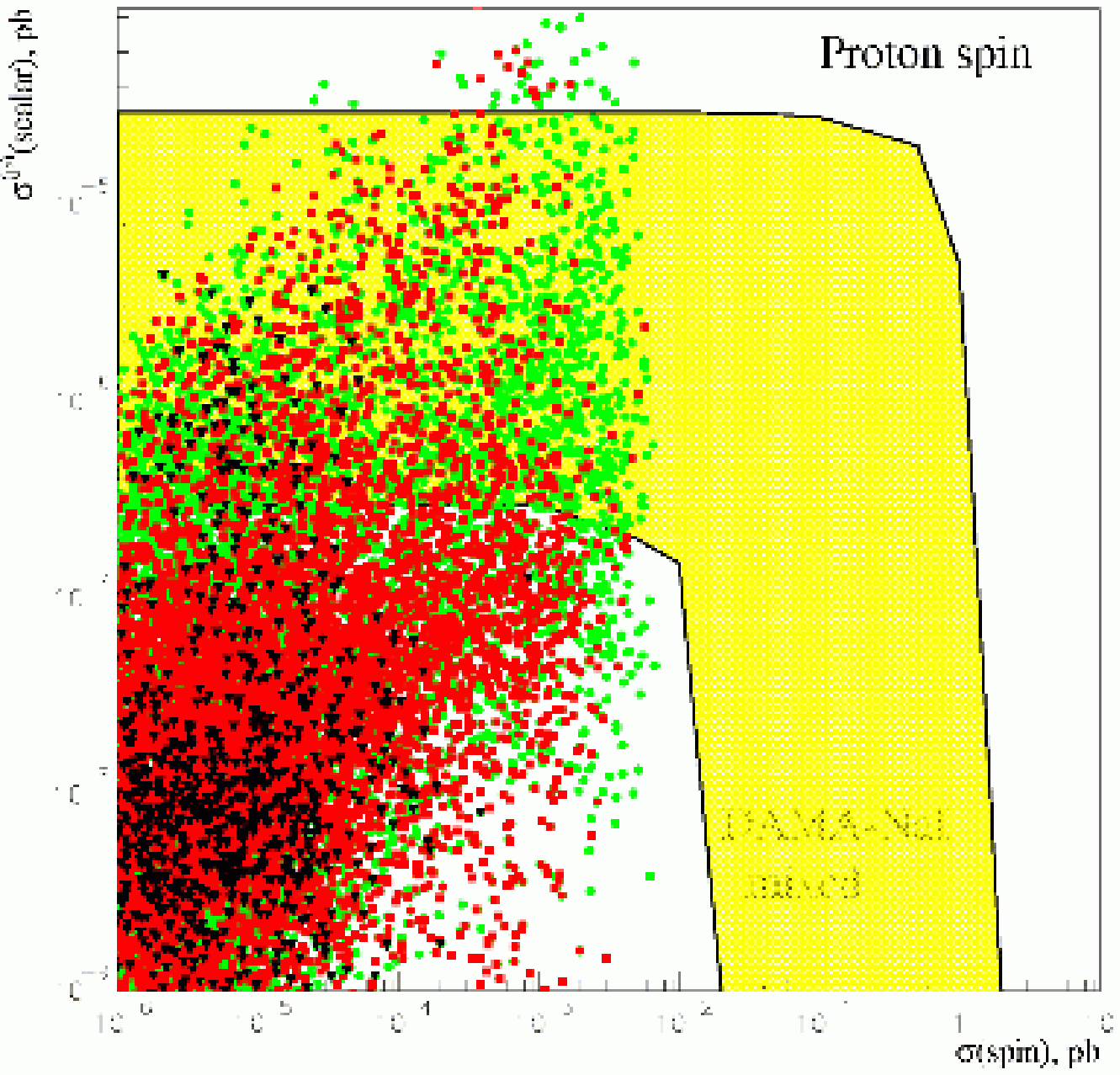}}
\put( 52,0){\includegraphics{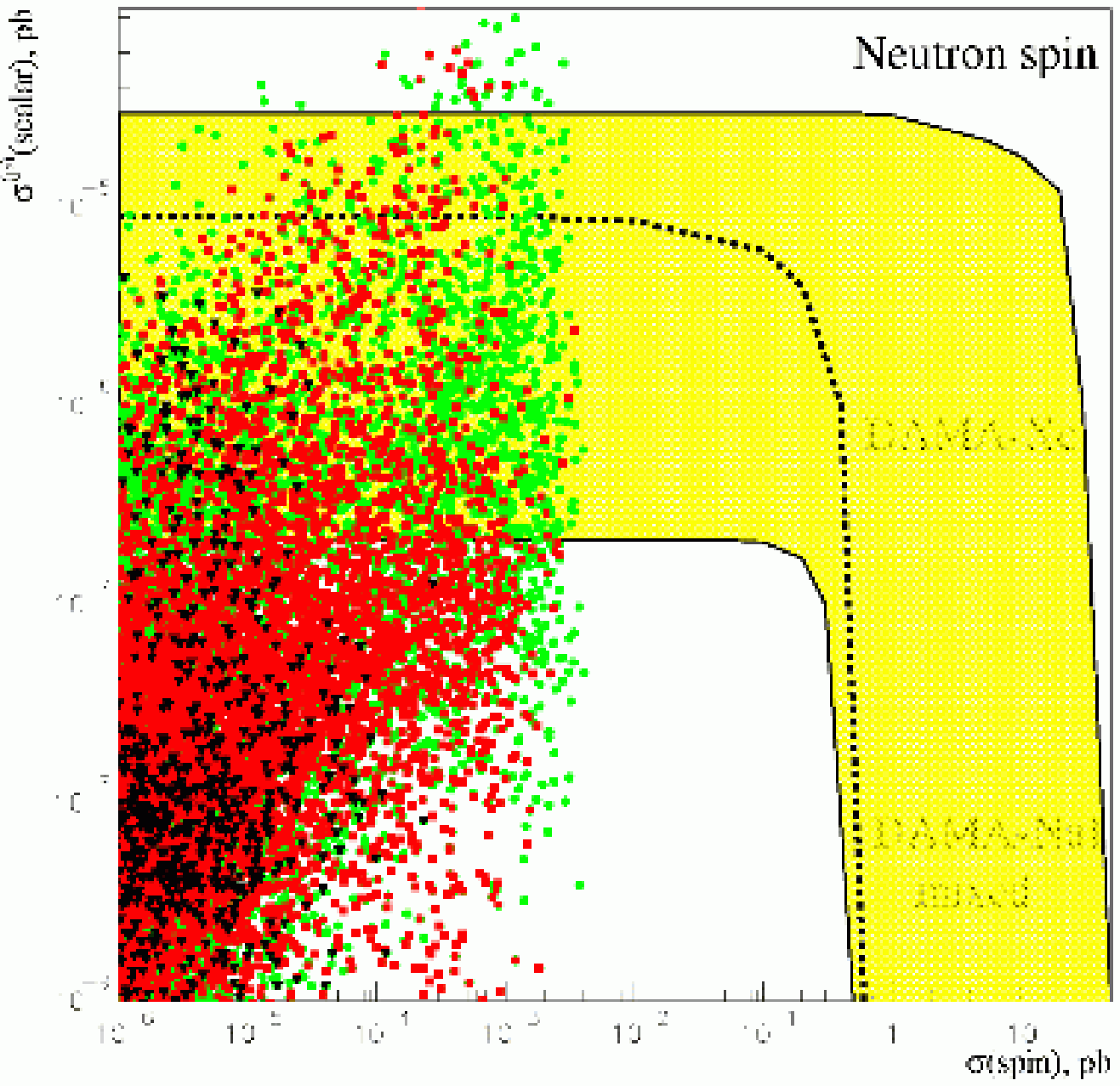}}
\end{picture}
\caption{The DAMA-NaI allowed region 
        from the WIMP annual modulation signature 
        in the ($\xi \sigma_{\rm SI}$, $\xi \sigma_{\rm SD}$) space 
        for $40<m^{}_{\rm WIMP}<110$~GeV
\cite{Bernabei:2003za,Bernabei:2001ve}.
        The left panel corresponds to the dominating (in $^{127}$I)  
        SD WIMP-proton coupling alone ($\theta$ = 0)  
        and the right panel corresponds to 
        the subdominating SD WIMP-neutron coupling alone 
        ($\theta$ = $\pi/2$). 
        The scatter plots give correlations
        between $\sigma^{p}_{{\rm SI}}$ and 
        $\sigma^{}_{{\rm SD}}$ in the effMSSM ($\xi=1$ is assumed) 
        for $m_\chi<200$~GeV 
\cite{Bednyakov:2004be}.
        In the right panel also the DAMA liquid xenon exclusion curve from 
\cite{Bernabei:2001ve} is given (dashed line).
From \cite{Bednyakov:2004be}.}
\label{Bernabei:2001ve:fig}
\end{figure}

In Fig.~\ref{Bernabei:2001ve:fig}
        the WIMP-nucleon spin and scalar mixed couplings 
        allowed by the annual modulation signature from 
        the 100-kg DAMA/NaI experiment 
        are shown inside the shaded regions. 
        The regions from 
\cite{Bernabei:2003za,Bernabei:2001ve} 
        in the 	($\sigma_{\rm SI}$, $\sigma_{\rm SD}$) 	space 
        for 40~GeV$<m^{}_{\rm WIMP}<$110~GeV cover
	spin-scalar mixing coupling for the proton ($\theta=0$ case of 
\cite{Bernabei:2003za,Bernabei:2001ve}, left panel) and 
	spin-scalar mixing coupling for the neutron 
	($\theta$ = $\pi/2$, right panel). 
	From nuclear physics one has for the proton spin dominated
        $^{23}$Na and $^{127}$I \
        $\frac{\langle {\bf S}_n\rangle}{\langle {\bf S}_p\rangle}< 0.1$ 
	and 
        $\frac{\langle {\bf S}_n\rangle}
	{\langle {\bf S}_p\rangle}< 0.02 \div 0.23$,
	respectively.
        For $\theta=0$ due to the p-oddness of the I target, 
	the DAMA WIMP-proton spin constraint is the most severe one
(see Fig.~\ref{Spin-p}).  
	In the right panel of  
Fig.~\ref{Bernabei:2001ve:fig}
	we also present the exclusion curve (dashed line) for the
	WIMP-proton spin coupling from the proton-odd isotope 
	$^{129}$Xe obtained under the mixed coupling assumptions  
\cite{Bernabei:2001ve} from the DAMA-LiXe (1998) experiment 
\cite{Bernabei:2002qg,Bernabei:1998ad,Bernabei:2002af}.
	For the DAMA NaI detector the 
	$\theta=\pi/2$ means no 
${\langle {\bf S}_p\rangle}$ contribution at all. 
	Therefore, in this case DAMA gives the 
        sub-dominant ${\langle {\bf S}_n\rangle}$ 
	contribution only,
        which could be compared further with the dominant 
        ${\langle {\bf S}_n\rangle}$ contribution in $^{73}$Ge.
  	The scatter plots in
Fig.~\ref{Bernabei:2001ve:fig} give 
	$\sigma^{p}_{{\rm SI}}$ as a function of 
	$\sigma^p_{{\rm SD}}$ (left panel) and  
	$\sigma^n_{{\rm SD}}$  (right panel)
	calculated in the effMSSM with parameters from 
Eq.~(\ref{Scanning})
	under the same constraints on the 
	relic neutralino density as in 
Figs.~\ref{CrossSections-vs-lsp}
\cite{Bednyakov:2004be}. 
	Filled circles (green) correspond to relic neutralino density 
	$0.0< \Omega_\chi h^2_0<1.0$,
	squares (red) correspond to sub-dominant relic neutralino
	contribution $0.002 < \Omega_\chi h^2_0<0.1$ 	
	and triangles (black) correspond to 
	the WMAP density constraint 
	$0.094 < \Omega_\chi h^2_0<0.129$
\cite{Spergel:2003cb,Bennett:2003bz}. 

	The constraints on the SUSY parameter space in the mixed coupling 
	framework in Fig.~\ref{Bernabei:2001ve:fig} 
	look, in general, 
	much stronger in comparison with the 
	traditional approach based on the one-coupling dominance.
        It follows from 
Fig.~\ref{Bernabei:2001ve:fig} 
        that when the LSP is the subdominant DM particle (squares in the
        figure), SD WIMP-proton and WIMP-neutron cross sections 
        at a level of $3\div5\cdot 10^{-3}$~pb are allowed,
        but the WMAP relic density constraint (triangles)
        together with the DAMA restrictions leaves only 
        $\sigma_{\rm SD}^{p,n}<3\cdot 10^{-5}$~pb
        without any visible reduction of allowed values for
        $\sigma^p_{\rm SI}$.
        In general, according to the DAMA restrictions, 
        very small SI cross sections are completely excluded, 
        only $\sigma^p_{\rm SI}> 3\div5\cdot 10^{-7}$~pb are allowed. 
	The SD cross section is not yet restricted at all.
	It is seen that 
        for the allowed values of the SI contribution  
        the SD DAMA sensitivity did not yet reach  
        the upper bound for the SD LSP-proton
        cross section of $5\cdot 10^{-2}$~pb 
	calculated for the nucleon spin structure from  
\cite{Ellis:2000ds}.

	In general, the famous DAMA ``conflict'' with the other 
	(negative) DM results 
	can be safely bypassed on the basis of the 
	above-mentioned mixed spin-scalar coupling approach, 
	where both SD and SI couplings are considered 
	simultaneously as non-negligible.

\section{The mixed coupling approach for the high-spin \boldmath 
$^{73}$G\lowercase{e}}
         In this section, 
	 on the basis of the high-spin $^{73}$Ge  
	 detector HDMS
\cite{Klapdor-Kleingrothaus:2002pg,Klapdor-Kleingrothaus:2000uh}, 
         the mixed spin-scalar coupling approach is used 
{\em to demonstrate how one can significantly improve}\   
         the quality of the exclusion curves
	 in comparison with the one-coupling dominance result of  
\cite{Klapdor-Kleingrothaus:2005rn}.

\begin{figure}[!ht] 
\begin{picture}(100,65) 
\put(-38,0){\includegraphics{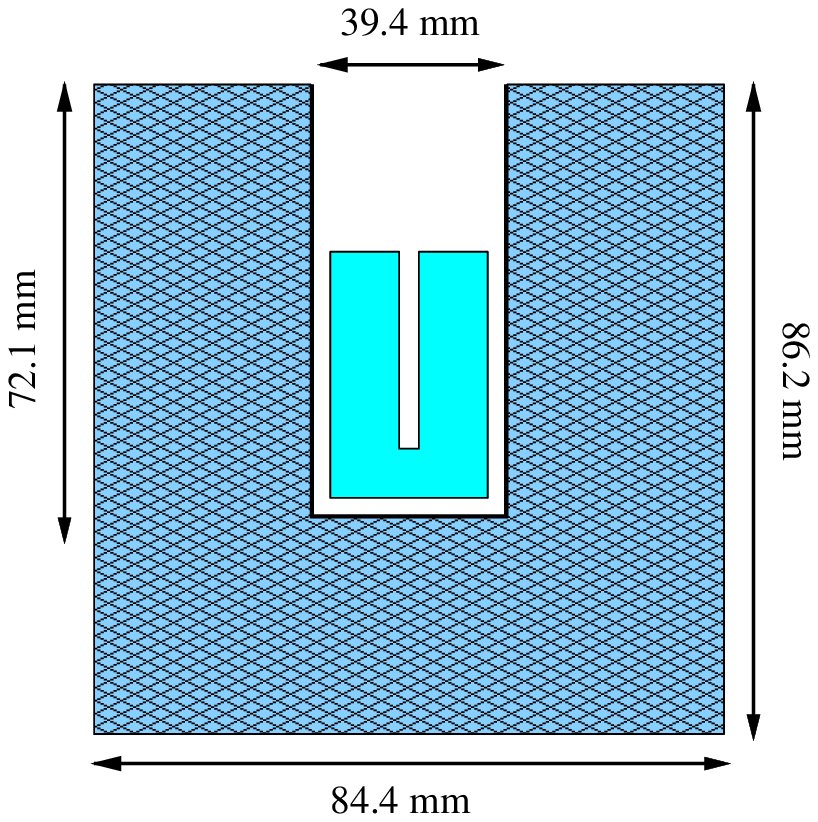}}
\put( 34,-38){\includegraphics{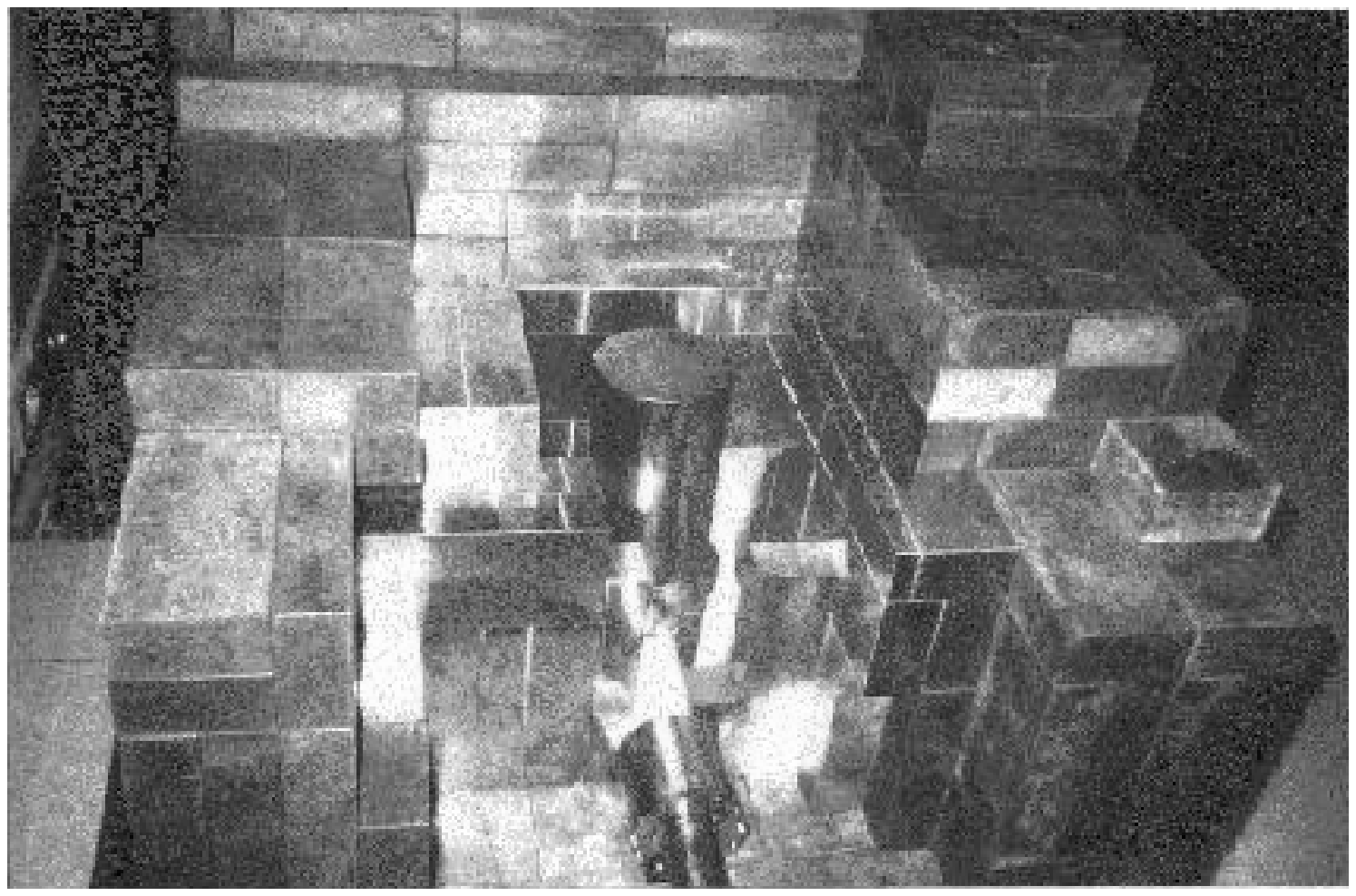}}
\end{picture}
\caption{Left: Schematic view of the HDMS detector configuration. 
         A small Ge crystal is surrounded by a well type Ge-crystal,
         the anti-coincidence between them is used to suppress background
         created by external photons. 
         Right: The HDMS detector in its open shield
         during the installation in the Gran Sasso Underground Laboratory.
         The inner shield is made of 10\,cm of electrolytic copper, 
         the outer one of 20\,cm of Boliden lead.
From~\cite{Klapdor-Kleingrothaus:2000cz,Klapdor-Kleingrothaus:2000uh}. 
        }
\label{HDMS:setup}\label{detindet}
\end{figure}

	The Heidelberg Dark Matter Search (HDMS) experiment
	used a special configuration of 
	two Ge detectors to efficiently reduce the background
(due to anti-coincidence of inner and outer detectors) 
\cite{Klapdor-Kleingrothaus:2005rn,%
Klapdor-Kleingrothaus:2002pg,Klapdor-Kleingrothaus:2000uh}.
        A small, p-type Ge crystal 
	(enriched by 86\% in $^{73}$Ge)
	is surrounded by a well-type natural 
        Ge crystal, both being mounted into a common cryostat 
        system (see left panel in  
Fig.~\ref{detindet} for a schematic view). 
     The HDMS with enriched $^{73}$Ge inner detector 
     was the first and till now unique setup with a high-spin (J=9/2) 
     Ge target isotope for direct DM search.
         The main idea of the new combined analysis relies 
	 on the unique possibility that two different isotope targets 
	 (from natural Ge and enriched $^{73}$Ge) were used as inner 
	 detector in the same HDMS setup under the same outer background
	 conditions of LNGS.

	In fact the first 
	simple estimation of the prospects
	for DM search and SUSY constraints with the high-spin $^{73}$Ge
	detector HDMS assuming mixing of WIMP-neutron spin and 
	WIMP-nucleon scalar couplings together with 
	available results from the 
	DAMA-NaI and LiXe experiments
\cite{Bernabei:2000qi,Bernabei:2003za,Bernabei:2003wy,%
      Bernabei:2002qg,Bernabei:1998ad,Bernabei:2002af}
        was performed in  
\cite{Bednyakov:2004be}.
	Furthermore, recently 
	in the mixed spin-scalar coupling approach
        the data from {\em both}\/ HDMS experiments
        with {\it natural} Ge and with {\it enriched} $^{73}$Ge were
        {\em simultaneously}\/ re-analyzed.
        This new analysis together with
        a new procedure for background identification and subtraction
        from the measured $^{73}$Ge spectrum allowed one 
	to obtain a significant (about one order of magnitude)
        improvement for the limits on the
        WIMP-neutron spin-dependent coupling.
	As a result the HDMS experiment is now giving
        the most sensitive limits on the WIMP-neutron spin coupling
        for WIMP masses larger than 60--65 GeV/$c^2$
\cite{Bednyakov:2007yf}. 

      The evaluation of the DM limits (exclusion curves) 
      on the WIMP-nucleon SD or SI cross section follows, in general
      (see, for example 
\cite{Klapdor-Kleingrothaus:2002pg,Baudis:2000ph,Baudis:1998hi}), 
      the conservative assumption that the whole experimentally measured 
      spectrum is saturated by       the WIMP-induced events. 
      Consequently, any excess events from the calculated
      spectrum above the relevant experimental spectra 
      in any energy interval are considered as 
      forbidden (at a given confidence level).
      In our case we assume that for any given WIMP mass $m_\chi$
      both $\sigma_\SI\equiv\sigma_\SI(m_\chi)$
      and  $\sigma_\SD\equiv\sigma_\SD(m_\chi)$
      WIMP-nucleon interaction cross sections are 
      excluded if 
\begin{equation}\label{fit-condition}
\frac{dR(\ER)}{d\ER} = 
          \kappa^{}_\SI(\ER,m_\chi)\,\sigma_\SI
         +\kappa^{}_\SD(\ER,m_\chi)\,\sigma_\SD 
         \ > \ 
         \frac{dR^{\rm data}}{d\ER}(\eth,\ER) 
	 \equiv R^{\rm data}(\eth,\ER), 
\end{equation}
      or both upper limit values for 
      $\sigma_\SI$ and $\sigma_\SD$ can be obtained
      as solutions of the following equation 
$$
           \kappa^{}_\SI(\ER,m_\chi)\,\sigma_\SI
          +\kappa^{}_\SD(\ER,m_\chi)\,\sigma_\SD 
         = R^{\rm data}(\eth,\ER)  
$$
      for all available recoil energies 
      $\ER$ above the threshold energy $\eth$.
    The notations are given by
Eqs.~(\ref{effectiveSD-cs-pn}), (\ref{effectiveSD-cs}),
(\ref{Definitions.diff.rate1}), and (\ref{structure}). 
	The sub-dominant contribution from 
	WIMP-proton spin coupling proportional 
	to $\langle {\bf S}^A_p\rangle$ can be safely neglected
	for $^{73}${Ge}. 
        The $^{73}$Ge isotope looks with a good accuracy 
        like an almost pure 
        neutron-odd group nucleus with 
        $\langle {\bf S}_{n}\rangle\! \gg\! \langle {\bf S}_{p}\rangle$
(Table~\ref{Nuclear.spin.main.table.71-95}).
        Therefore in our consideration 
$\sigma^{}_{\SD} \equiv \sigma^{n}_{\SD}$ and $\cos\theta=0$.
        For the WIMP mass density in our Galaxy 
        the value $\rho_\chi=0.3$~GeV$/$cm$^3$ is used. 
     
\begin{figure}[h!] 
\begin{picture}(100,60)
\put(-35,0){\includegraphics{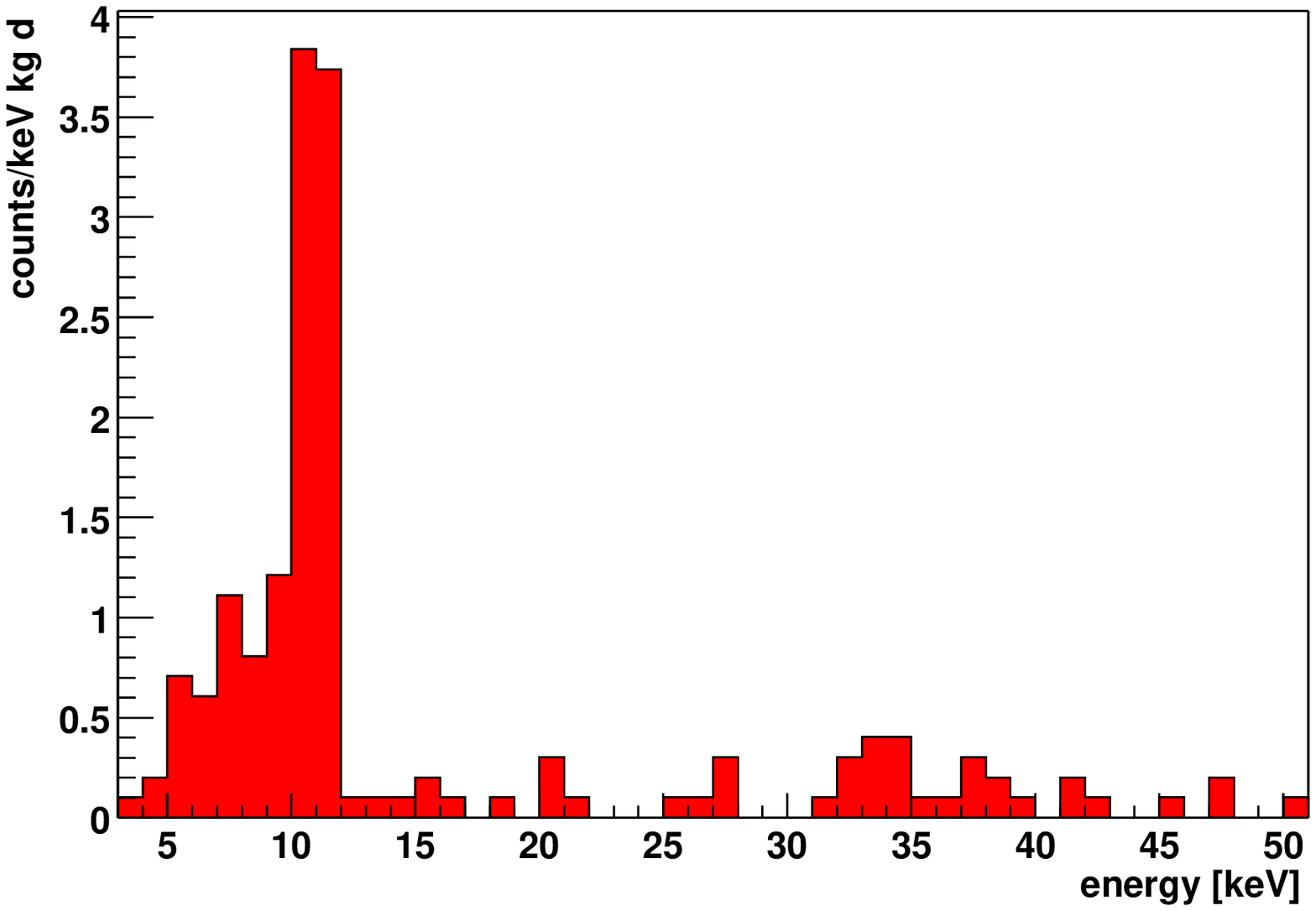}}
\put( 51,0){\includegraphics{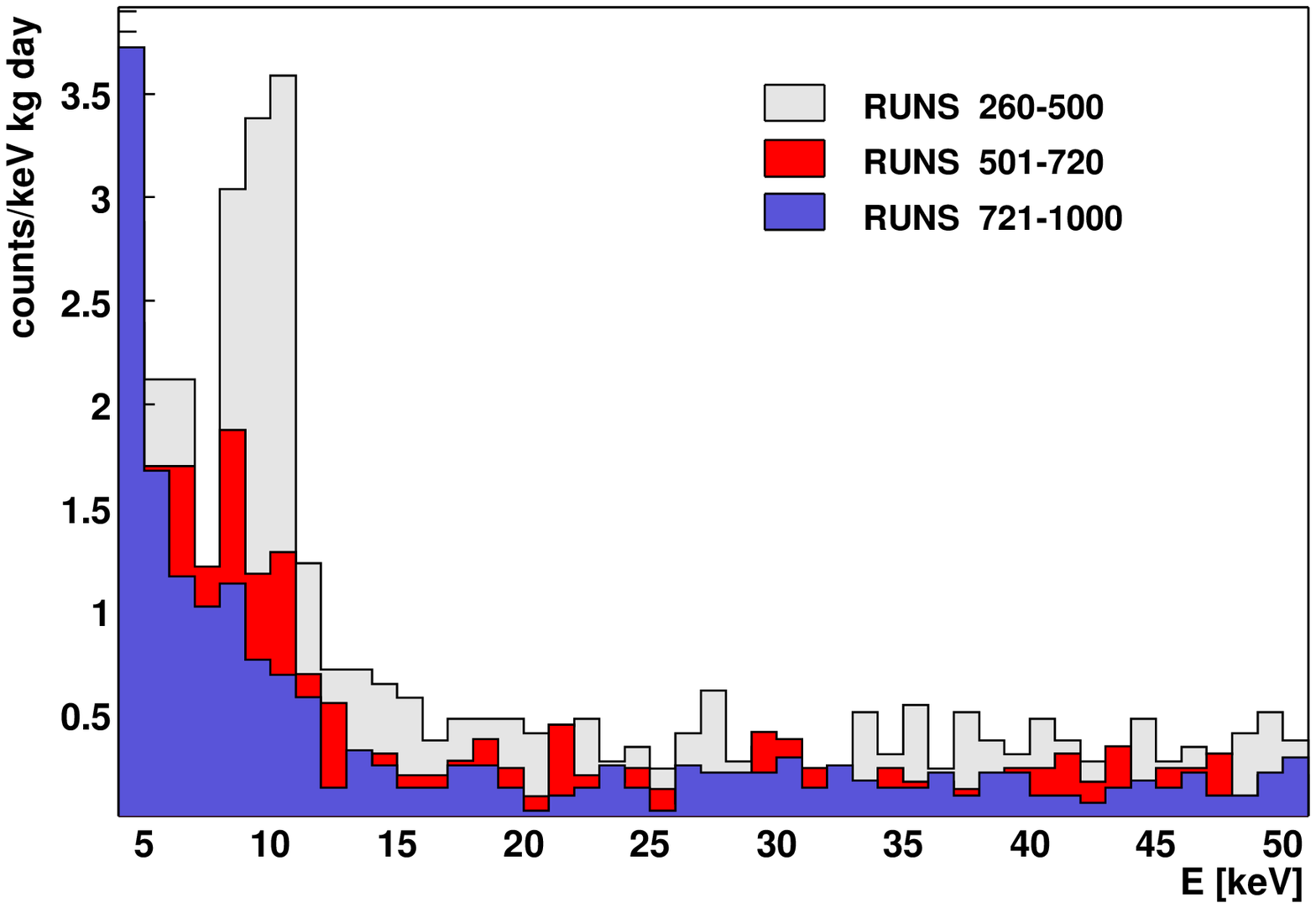}}
\end{picture}
\caption{HDMS spectra used in the analysis. 
        Left panel: 
        Spectrum obtained in the 
        first prototype phase of the HDMS experiment with the 
        inner detector from natural Ge (exposure 9.9 kg d)
\cite{Baudis:2000ph}.
        Right panel: 
        Spectrum obtained from the full HDMS
         setup with the inner detector from enriched $^{73}$Ge
        (exposure 85.5 kg d) 
\cite{Klapdor-Kleingrothaus:2005rn}. 
         It is separated in 3 sets of runs of 
         30.9, 29.5 and 27.6 kg d, respectively.
         The blue histogram corresponds to the latest
         (most clean) runs 721--1000.
	From 
\cite{Klapdor-Kleingrothaus:2005rn}. 
}
\label{HDMS:spectra}\label{HDMS:73-spectrum}
\end{figure}

        To find both $\sigma^{}_\SI$ and $\sigma^{}_\SD$, 
	for any given $m_\chi$, in accordance with  
Eq.~(\ref{fit-condition}) 
        the following functional
\begin{equation}
\label{chi-2}
\chi^2(m_\chi,\eth) = \sum^{{\rm spectra}}_{j}
                 \sum^{{\rm bin}}_{i}
\frac{\left(R^j(\eth,E_i)
 -\kappa^{}_\SI(E_i,m_\chi)\,\sigma_\SI
 -\kappa^{j}_\SD(E_i,m_\chi)\,\sigma_\SD 
   \right)^2}{(\Delta R^j(E_i))^2}  
\end{equation}
      can be numerically minimized, 
     where $R^j(E_i)$ and $\Delta R^j(E_i)$ are
     measured rate and its error (in counts/day/kg/keV) 
     in $i$-th energy bin for $j$-th used spectrum 
     ($j=1,2$ for natural Ge and Ge-73).
     The two main used spectra are given in
Fig.~\ref{HDMS:spectra}.
	  Only the ``cleanest'' background spectrum 
          with the $^{73}$Ge target collected in the latest runs
          of the experiment (with numbers 721--1000) 
	  was used in the analysis (blue color histogram in 
	  the right panel of Fig.~\ref{HDMS:73-spectrum}).
         For both spectra 
	 the visual energy threshold 
	 $\eth=4$~keV is used. 

         Comparing both the 'most accurate' (blue, runs 721-1000 in 
Fig.~\ref{HDMS:73-spectrum}) HDMS $^{73}$Ge-spectrum
         and the natural Ge spectrum 
	 one can see obviously  some non-vanishing {\em extra} 
         background contribution in the first spectrum 
         relative to the second one.
         In general, such a possibility is not new. 
          The improvement in the exclusion curves by taking into account known 
          sources of background during DM searches with Ge detectors
          was demonstrated, for example, in 
\cite{Garcia:1995ez} and further discussed in 
\cite{Baudis:1998wu}.
         Therefore for this HDMS $^{73}$Ge-spectrum 
         we allow the possibility to fit simultaneously
         with the SD and SI cross sections some constant 
	 (as function of the recoil energy)
         background contribution, too.
         The effect of this extra background 
         contribution is discussed later on.

	  For a semiconductor germanium detector
	  one has to take into account the ionization efficiency.
	  For the HDMS Ge setup in the (visual) energy interval
	  $4 < \ER <50~{\rm keV}$ a simple relation between the 
	  visible recoil energy 
	  and the real recoil energy --- 
	  $E^{}_{\rm vis} \equiv \ER = 0.14 E^{1.19}_{\rm real~recoil}
     \approx Q E^{}_{\rm real~recoil}$ ---
     can be used 
     with  $Q=0.33$ being the quenching factor for Ge
\cite{Klapdor-Kleingrothaus:2002pg,Baudis:2000ph,Baudis:1998hi,%
Klapdor-Kleingrothaus:2005rn}. 

	 One can note that for any WIMP mass, $m_\chi$, and 
	 any target mass, $M_A$, 
	 due to kinematics one has not to
	 expect any WIMP-induced event at all with
\begin{equation}\label{Maximal-Recoil}
\ER > {\ER}^{\max} =
  Q\frac{2 v_{\max}^2 M_A m_\chi^2}{\left ( M_A + m_\chi \right)^2}
= Q \frac{2 v_{\max}^2}{M_A} \mu^2_A,
\qquad 
v_{\max} \approx \vesc.
\end{equation}
     For example, 
     ${\ER}^{\max} = 4~(50)$~keV, for a Ge detector 
     and $m^{}_\chi = 12~(77)$~GeV/$c^2$.
     It is clear from
Eq.~(\ref{Maximal-Recoil}) that for fixed $v_{\max}$, 
     $M_A$ and the detector energy threshold $\eth$
     there are undetectable WIMPs (with rather light $m_\chi$) if 
     the maximal recoil energy, they can produce,  
     is smaller than the threshold:
     ${\ER}^{\max} < \epsilon$.
     Therefore one has two restrictions for 
     the theoretical event rate 
     as function of the WIMP mass $m_\chi$:
\begin{equation}
\label{WIMP-mass-restrictions}
R(\ER>{\ER}^{\max}(m_\chi)) \equiv 0 
\qquad {\rm and} \qquad
R ({\ER}^{\max}(m_\chi)<\eth) \equiv 0.
\end{equation}
      The first one allows 
      background (for the WIMP-nucleus signal) estimation 
      in the $^{73}$Ge measured spectrum, 
      which could lead to a remarkable improvement of the deduced 
      exclusion curves.

         Now we turn to our main analysis 
of both HDMS spectra 
         in the mixed spin-scalar coupling approach and extract limits for 
         both cross sections 
         $\sigma^{}_\SI$ and $\sigma^{}_\SD$ simultaneously 
         using formulas 
(\ref{fit-condition})--(\ref{chi-2}). 
	 To obtain from the available data the 
	 most accurate exclusion curve 
	 one can use two minimization approaches.
	 The first (main) approach relied on indeed direct simultaneous 
	 determination of the SD and SI WIMP-nucleon limits
	 for a given WIMP mass $m_\chi$
	 (exclusion curves for 
	 $\sigma^{}_{\SD}(m_\chi)$ and 
	 $\sigma^{}_{\SI}(m_\chi)$) by means of 
	 minimization  of the discrepancy 
	 between our calculated estimations of the expected rates and  
	 both above-mentioned experimental spectra. 
	 In our second (auxiliary) approach assuming 
	 SI coupling dominance ($\sigma^{}_{\SD}=0$) we
	 first extracted only the SI WIMP-nucleon limit 
	 $\sigma^{}_{\SI}(m_\chi)$
	 from the natural Ge spectrum (left panel in 
Fig.~\ref{HDMS:spectra}).
          Next, for each WIMP mass and above defined 
	  $\sigma^{}_{\SI}(m_\chi)$ 
	  we extracted only the SD WIMP-nucleon limit 
	  $\sigma^{}_{\SI}(m_\chi)$ 
	  from the cleanest background spectrum 
          with $^{73}$Ge 
         (Runs 721--1000,  27.6 kg d, blue spectrum in right panel of 
Fig.~\ref{HDMS:spectra}).
          In both cases the obtained results are rather similar.

	 Furthermore trying to improve the quality of the exclusion curves  
	 one can 
	 use a sliding variable energy window to check the excess events 
	 above the experimental spectrum 
	 (in these energy window intervals)
	 as used in previous papers 
\cite{Klapdor-Kleingrothaus:2005rn,Klapdor-Kleingrothaus:2002pg,Klapdor-Kleingrothaus:2000uh,Baudis:2000ph}.
       The minimum among the cross section values obtained via the multiple 
       fits is taken as the cross section for the corresponding WIMP mass.
       We used 5 keV minimal width of this energy window as in 
\cite{Baudis:1998hi,Baudis:2000ph} and a 10-keV window as well.

\begin{figure}[h!] 
\begin{picture}(100,105)
\put(-37,-8){\includegraphics{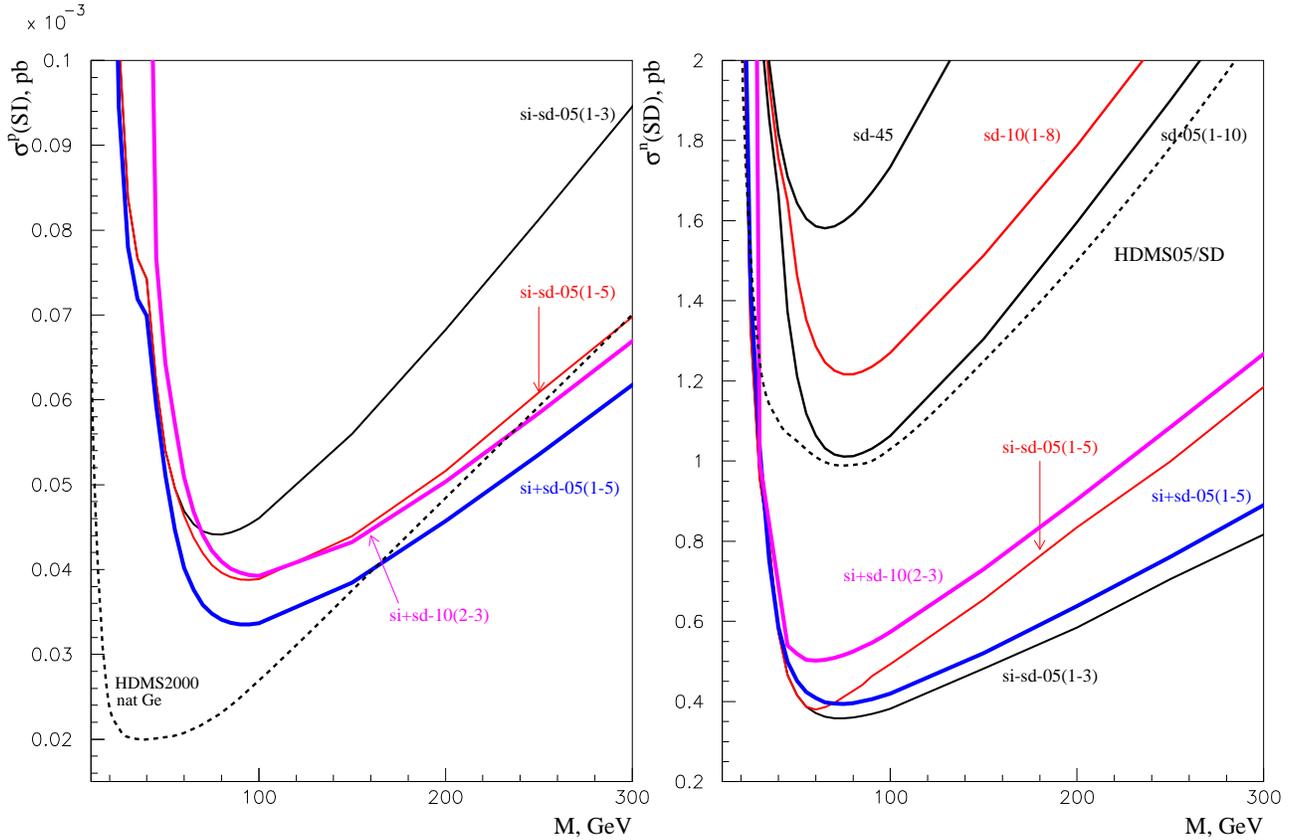}}
\end{picture}
\caption{
        Spin-independent $\sigma_\SI$ (left panel) and 
	spin-dependent $\sigma^{}_{\SD}$ 
	(right panel)
	cross section upper limits 
	as function of WIMP mass 
	(exclusion curves) obtained  
	from {\em simultaneous}\/ analysis 
        of both HDMS Ge spectra under different 
	minimization conditions.
	 The label ''si-sd-05(1-5)'' means that 
	 {\em only}\/ the spectrum from natural Ge (left panel, 
Fig.~\ref{HDMS:spectra})
        is used first to extract a SI limit (auxiliary approach) 
	using 5-keV sliding window and taking into account only
the first 5 lowest energy windows to obtain the limits.
        Curves (blue) labeled with 
	``sd+si-05(1-5)'' are obtained from indeed {\em simultaneous}\ 
	minimization of both
	natural Ge and the best (right panel, 
Fig.~\ref{HDMS:spectra})
        enriched $^{73}$Ge spectra using 
	the 5-keV sliding window and 
	only the first 5 lowest energy windows.
	The label ''si+sd-10(2-3)'' denotes the same procedure, 
	but with 10-keV sliding window and with 
	the 2nd and the 3rd lowest energy windows.
	The other labels have analogical meaning.
         Dashed lines correspond to some other analyses 
         and are given for comparison 
         (``HDMS2000, nat Ge'' from 
\cite{Baudis:2000ph}, ``HDMS05/SD'' from 
\cite{Klapdor-Kleingrothaus:2005rn}). 
	The thin exclusion curves 
	``sd-45'', ''sd-10(1-8)'' and ''sd-05(1-10)'' 
	are obtained from the traditional 
	one-coupling dominance fit of the 
        enriched $^{73}$Ge spectrum {\em only}\/ and are given	to 
	illustrate the role of the sliding window width
	as well as the consistency with the previous result of
\cite{Klapdor-Kleingrothaus:2005rn}. 
}
\label{mixed-couplings-1}
\end{figure}

         First, the possible 
	 improvements of the exclusion curves 
	 due to variation of  minimization conditions, used 
	 in the data analysis, were studied. 
	 The relevant 
         exclusion curves obtained from the {\em simultaneous}\/
	 analysis 	 of {\em both}\/ HDMS spectra 
	 within the mixed spin-scalar coupling approach 
	 are given in 
Fig.~\ref{mixed-couplings-1} as function of the WIMP mass.

        The left panel of  
Fig.~\ref{mixed-couplings-1} shows the upper limits for the 
        spin-independent WIMP-nucleon cross section
        $\sigma_\SI$ as function of WIMP mass
	obtained under different minimization conditions.
	The right panel of 
 Fig.~\ref{mixed-couplings-1} shows the upper limits for the 
        spin-dependent WIMP-neutron cross section
        $\sigma_\SD$ (in our approximation 
        $\sigma^{}_{\SD} \equiv \sigma^{n}_{\SD}$)
	as function of WIMP mass which
        correspond to the above-mentioned
	$\sigma_\SI$ limits from the left panel.
	For example, 
	the (red) exclusion curve labeled with ''si-sd-05(1-5)'' 
	presents in the left panel 
	the SI limits $\sigma^{}_{\SI}(m_\chi)$
	extracted 
	 {\em only}\/ from the natural Ge spectrum (left panel, 
Fig.~\ref{HDMS:spectra}) 
	with 5-keV sliding windows within 
	only the first 5 lowest energy windows
	and under the assumption $\sigma_\SD=0$.
	The relevant 
	(red) ''si-sd-05(1-5)'' exclusion curve 
	in the right panel shows {\em correlated}\/ 
	SD limits $\sigma^{}_{\SD}(m_\chi)$
	extracted from the best enriched $^{73}$Ge spectrum  
	(Runs 721--1000, blue spectrum, right panel, 
Fig.~\ref{HDMS:spectra})
	when 	$\sigma^{}_{\SI} = \sigma^{}_{\SI}(m_\chi)\neq 0$
	from the left panel.
        The (thick blue) curves labeled with 
	``sd+si-05(1-5)'' are obtained from 
	simultaneous minimization of both
	above-mentioned spectra from 
Fig.~\ref{HDMS:spectra} 
	using the 5-keV sliding window and 
	the first 5 lowest energy windows.
	The other exclusion curves 
	(with labels ``sd+si-10(2-3)'' and ``sd-si-05(1-3)'')
	are given in 
Fig.~\ref{mixed-couplings-1} to illustrate the
         exclusion curve dependence on 
	 the width of the 
	 sliding window and optimal (or non-optimal) choice
	 of minimization regions.

        The dashed lines correspond to some other analyses 
         and are given for comparison 
         (``HDMS2000, nat Ge'' from 
\cite{Baudis:2000ph}, ``HDMS05/SD'' from 
\cite{Klapdor-Kleingrothaus:2005rn}). 
	The thin exclusion curves 
	``sd-45'', ''sd-10(1-8)'' and 
	''sd-05(1-10)'' 
	are obtained from the traditional 
	one-coupling dominance fit of the enriched $^{73}$Ge spectrum 
	only and are given also to 
	illustrate the role of the width of the sliding window
	as well as the consistency with the previous result of
\cite{Klapdor-Kleingrothaus:2005rn}. 
         The small discrepancy between the ``HDMS05/SD'' and 
	 ''sd-05(1-10)''  curves for low WIMP masses is mainly 
	 due to our restrictions 
(\ref{WIMP-mass-restrictions}).

	The pair of SI and SD exclusion curves with label ``sd+si-05(1-5)''
	corresponds to the optimal parameters of the fitting procedure
	and presents the best correlated exclusion curves
	obtained from 
	simultaneous minimization of both
	spectra from natural Ge spectrum 
        and the ``cleanest'' enriched $^{73}$Ge spectrum. 
        The pair has the best SI exclusion curve (left panel) simultaneously
	with almost the best SD exclusion curve (right panel). 
	The visible difference between the best SI 
	curve ``sd+si-05(1-5)'' and the ``HDMS2000, nat Ge'' curve from 
\cite{Baudis:2000ph} is due to restrictions
(\ref{WIMP-mass-restrictions}) and another
        energy threshold (see below) used in our analysis. 

         Therefore, from the right panel of 
Fig.~\ref{mixed-couplings-1} 
         one can conclude that the most sensitive 
         exclusion curves for the WIMP-{\em neutron}\/ spin interaction
         (``si-sd-05(1-3)'', black and ``si+sd-05(1-5)'', blue)
         improve the relevant one-coupling dominance 
         limit of 	 ``HDMS05/SD'' 
\cite{Klapdor-Kleingrothaus:2005rn} 
         within a factor of 2--3 depending on the WIMP mass.
	 This is a clear result of the mixed spin-scalar approach.

         Now we consider two other possibilities to improve the
	 quality of the exclusion curves extracted from 
	 both Ge spectra within the mixed coupling scheme.
	 The first one is a lower recoil (visible) energy 
	 threshold for the natural Ge detector of HDMS setup.  
	 The second one 
	 is a new procedure of background subtraction 
	 from the {\em measured spectrum}\/ of the $^{73}$Ge isotope. 
	 This procedure {\em strongly}\/ relies on the existence of 
	 a really measured spectrum.

\begin{figure}[t!] 
\begin{picture}(100,100)
\put(-38,-8){\includegraphics{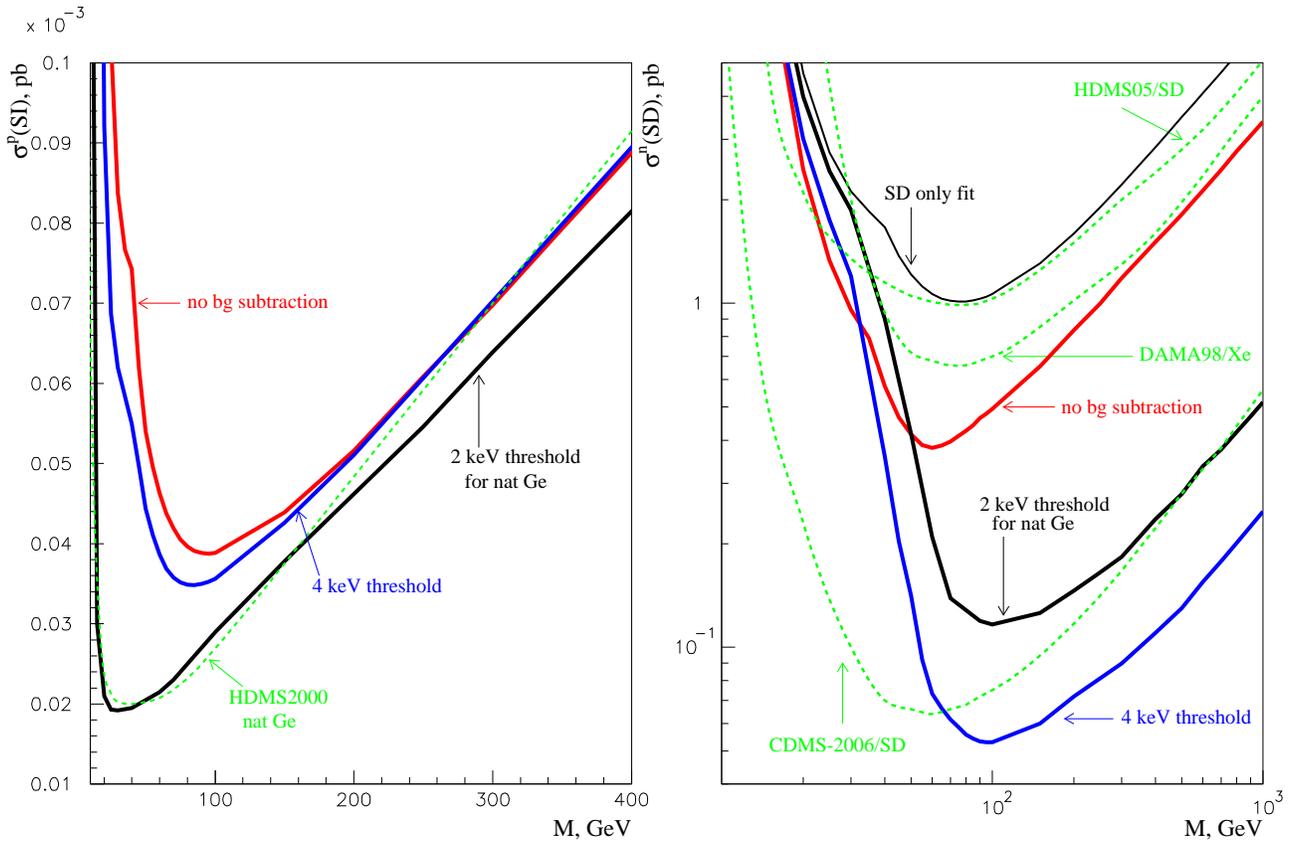}}
\end{picture}
\caption{Exclusion curves from {\em simultaneous}\/ fit 
         of the data from both HDMS setups.   
        Left panel: 
        Spin-independent cross section upper limits
        $\sigma_\SI$ in pb as function of WIMP mass in GeV. 
        Right panel: 
        Spin-dependent cross section upper limits $\sigma_\SD$ 
	as function of WIMP mass 
        which 
	correspond to the $\sigma_\SI$ limits from the left panel.
	 The label ``no bg subtraction'' 
	 shows the best exclusion curves (in red) 
	 obtained from simultaneous minimization of both
	 natural Ge (left panel, 
Fig.~\ref{HDMS:spectra})
        and the best (Runs 721--1000, blue spectrum, right panel, 
Fig.~\ref{HDMS:spectra})
        enriched $^{73}$Ge spectra within 
	mixed spin-scalar coupling approach 
	{\em without}\ any background subtraction from the
	$^{73}$Ge spectra.
        Curves (in blue) labeled with 
	``4 keV threshold'' are from the same 
	simultaneous minimization of both
	natural Ge  and the best enriched $^{73}$Ge spectra 
	but with extra background of 0.11 events/kg/day 
	extracted from the $^{73}$Ge  spectrum.  
	The curves (in black) labeled with 
	``2 keV threshold for nat Ge'' 
	are from the same procedure as above, but
	when threshold for the natural Ge spectrum is equal to 
	2 keV. The curve reproduces the best HDMS SI limits from 
\cite{Baudis:2000ph}
        given here as ``HDMS2000, nat Ge'' (dashed green).
	  The thin (black) exclusion curve 
	``SD only fit'' from the traditional 
	  one-coupling dominance fit of the 
          enriched $^{73}$Ge spectrum only 
	  is given for comparison with previous analysis of 
\cite{Klapdor-Kleingrothaus:2005rn} labeled with ``HDMS05/SD''.
	  The best exclusion curve for WIMP-neutron spin coupling 
	  from CDMS collaboration 
\cite{Akerib:2005za} 
	  is labeled with ``CDMS-2006/SD'' (dash green).
          The last dashed (green) line ``DAMA98/Xe'' 
	  corresponds to DAMA results from 
\cite{Bernabei:1998ad}.
           Another comparative result from ZEPLIN-I 
\cite{Alner:2005pa} (not shown)
	   is located above the CDMS curve nearby the black one. 
}
\label{mixed-couplings}
\end{figure} 

        In
Figure~\ref{mixed-couplings}  
         the thin (black) exclusion curve labeled with 
	 ``SD only fit'' 
	 is the result of the 
	one coupling dominance analysis of the 
        enriched $^{73}$Ge spectrum only. 
	It repeats the relevant curve (labeled with ``sd-05(1-10)'') from 
Fig.~\ref{mixed-couplings-1}.	  
	As mentioned before, 
	the curve is consistent with the previous analysis of 
\cite{Klapdor-Kleingrothaus:2005rn}, given here with 
	curve ``HDMS05/SD''. 
	
	 The curves labeled with ``no bg subtraction''
	 repeat here the best exclusion curves 
(``si+sd-05(1-5)'' in Fig.~\ref{mixed-couplings-1})
	 obtained 
	 within the mixed spin-scalar coupling approach 
         from simultaneous analysis of 
	 both natural Ge and the best enriched $^{73}$Ge  spectra.
	 The same recoil energy threshold of {\em 4 keV}\/ 
	 was taken for both Ge spectra.
	 This threshold corresponds to the real threshold of the
	 HDMS final setup with enriched $^{73}$Ge
\cite{Klapdor-Kleingrothaus:2002pg}.

         We reproduce
in Fig.~\ref{mixed-couplings}  both exclusion curves 
         ``SD only fit''  and ``no bg subtraction''
(from Fig.~\ref{mixed-couplings-1}) for our further consideration 
	 and with the aim to
	 clearly demonstrate again that 
	 the most sensitive HDMS exclusion curve 
	 (``no bg subtraction'')
	 for the WIMP-{\em neutron}\/ spin interaction
         improves the relevant one-coupling dominance 
         limit of  ``HDMS05/SD'' 
\cite{Klapdor-Kleingrothaus:2005rn}
	and the ``SD only fit'' curve
         within a factor of 2--3. 

	 We stress that this ``no bg subtraction''  curve is 
        {\em obtained 
	  from the raw HDMS data without any active or passive 
	  background substraction}.


 	The visible difference at low WIMP masses between SI exclusion 
	curve ``no bg subtraction'' and the ``HDMS2000, nat Ge'' 
	curve from 
\cite{Baudis:2000ph} (in the left panel of 
 Fig.~\ref{mixed-couplings})        is mainly due to  
        the lower recoil energy threshold of 2 keV used in
\cite{Baudis:2000ph} for the natural Ge detector. 
         The curve ``2 keV threshold for nat Ge''
(black in the left panel of Fig.~\ref{mixed-couplings}) 
	 obtained indeed with a 
	 2 keV energy threshold for the spectrum of natural Ge 
	 proves the reason of the difference. 
 
	Furthermore, the real measured 
	spectrum of enriched $^{73}$Ge 
	(Runs 721--1000, blue spectrum, right panel,
Fig.~\ref{HDMS:spectra}) 
        and the first relation  
(\ref{WIMP-mass-restrictions})
	 allow one to estimate some number of counts in the spectrum 
	 which can not be produced by means of 
	 any WIMP-nucleus interaction.
	 In accordance with 
(\ref{WIMP-mass-restrictions})
	 for any $m_\chi$, fixed $v^{}_{\max}$ and $M_A$
	 there is a maximal recoil energy ${\ER}^{\max}(m_\chi)$
(\ref{Maximal-Recoil}) 
	 for which WIMP-nucleus interactions are
	 unable to produce any signal if $\ER>{\ER}^{\max}(m_\chi)$
	 (i.e. when the measured recoil energy is larger than the 
	 maximally
	 possible recoil energy for a given WIMP mass).
	 Therefore, for fixed $m_\chi$ 
	 the measured recoil spectrum in the region 
	 $\ER>{\ER}^{\max}(m_\chi)$ is directly 
	 some background which can be approximated, for example, 
	 as a constant function of the recoil energy,  
	 independent of $m_\chi$.
	 One can estimate these background 
	 constants for each allowed $m_\chi$ 
	 (still ${\ER}^{\max}(m_\chi)<50$~keV)
	 and assume the minimal of these constants
	 (0.11 events/kg/day/keV) to be the mean background
	 for all measured ${\ER}$ and all $m_\chi$. 
	 Physically this extra background is completely independent on
	 any WIMPs, therefore being estimated
	 for rather small $m_\chi< 100$~GeV, 
	 it can be used for all WIMP masses as well.

	 Therefore, with common energy threshold of 4 keV,
	 the simultaneous minimization of both
	 natural Ge and the best $^{73}$Ge spectra
	 {\em with}\ 
	 the above-mentioned extra 
	 background of 0.11 events/kg/day 
	 has supplied us with 
	 the pair of SD and SI exclusion curves, 
	 labeled with ``4 keV threshold'' in 
 Fig.~\ref{mixed-couplings}.
         This SD curve improves (at least for 
	 $m_\chi> 60$~GeV) currently the 
	 best exclusion curve (labeled with ``CDMS-2006/SD'', dash green)
	 for the WIMP-neutron spin coupling 
	 from the CDMS collaboration 
\cite{Akerib:2005za}. 
        The other dashed lines correspond to some other analyses 
        and are given for comparison 
        (``HDMS2000, nat Ge'' from 
\cite{Baudis:2000ph}, ``HDMS05/SD'' from 
\cite{Klapdor-Kleingrothaus:2005rn}, and ``DAMA98/Xe'' from 
\cite{Bernabei:1998ad}).

	The curves (in black, right  panel of 
 Fig.~\ref{mixed-couplings}) 
         labeled with 
	``2 keV threshold for nat Ge'' 
	are from the same fit procedure with the extra background, 
	but when the threshold for the natural Ge 
	spectrum is equal to 2 keV. 
	In this case, as mentioned above,
	one reproduces the best HDMS SI limits 
	``HDMS2000, nat Ge'' (dashed green) from 
\cite{Baudis:2000ph}.
	 As it is seen, 
         the background subtraction from the $^{73}$Ge spectrum
	 only very weakly affects 
	 the correspondent SI curves in 
 Fig.~\ref{mixed-couplings} (left panel).

\begin{figure}[!h] 
\begin{picture}(100,110)
\put(-5,-5){\includegraphics{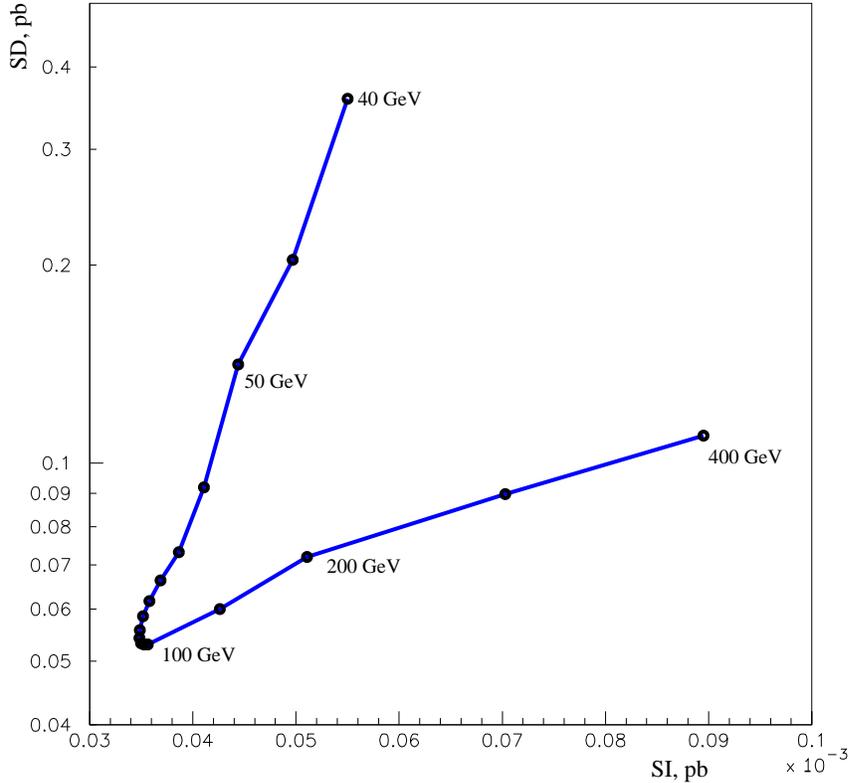}}
\end{picture}
\caption{Correlated spin-dependent cross section upper limit
         $\sigma_\SD$ and 
         spin-independent cross section upper limit
         $\sigma_\SI$ 
         obtained from the 
         simultaneous analysis of the data from both HDMS setups.
         Points running over the
         curves mark the relevant WIMP mass values.
         For example, the point marked with label ``200 GeV''
         shows the best simultaneous upper limits for
         SD and SI WIMP-nucleon interaction cross sections
	 for WIMP mass 200~GeV$/c^2$. 
         For the first time similar plots were given in 
\cite{Bernabei:2001ve}. 
From \cite{Bednyakov:2007yf}. 
}
\label{SD-vs-SI}
\end{figure}

\smallskip
        The main results of the analysis performed  
        in the mixed spin-scalar coupling approach 
        are the (correlated) limits for the cross sections 
        $\sigma^{}_\SI$ and $\sigma^{}_\SD$. 
	Indeed,
        despite the traditional form of presentation
        of the SD and SI exclusion curves in 
Fig.~\ref{mixed-couplings} as function of WIMP mass   
        one should keep in mind that these
        $\sigma^{}_{\SI}$ and $\sigma^{n}_{\SD}$
        constraints for fixed WIMP mass are
        strongly correlated.
        This correlation is presented explicitly in 
Fig.~\ref{SD-vs-SI},         
        where the dependence on WIMP mass is given 
        indirectly by means of the points running over the
        curves and marked with relevant WIMP mass values.
        For example, the point marked with 
	label ``200~GeV''
        shows the simultaneous upper limits for
        SI and SD WIMP-nucleon interaction cross sections
        $\sigma^{}_{\SI}$ and $\sigma^{n}_{\SD}$
        for $m_\chi = 200$~GeV$/c^2$. 
        This, in principle, gives one 
        a new requirement (for a SUSY-like theory)  that
        for any fixed WIMP mass $m_\chi$ one should have 
        $\sigma^{}_\SI({\rm theor.})
        \le \sigma^{}_\SI({\rm fitted})$ 
        and 
        $\sigma^{n}_\SD({\rm theor.})
        \le \sigma^{n}_\SD({\rm fitted})$ 
        {\em simultaneously}.

\smallskip
      For the sake of completeness one can compare the limits obtained 
      (in the spin-scalar mixed coupling approach) from the HDMS 
      experiments on the SD and SI WIMP-nucleon interaction  
      with the relevant constraints extracted by the DAMA 
      collaboration from measurement of the
      annual signal modulation with NaI target
\cite{Bernabei:2003za}.
        Following the DAMA positive evidence one can  
        accept that the most prefered interval of the WIMP mass is 
$40~{\rm GeV}/c^2 < m^{}_{\rm WIMP} < 110~{\rm GeV}/c^2.
$
        Other consequences of the fact one can find in
\cite{Bednyakov:2005qp,Bednyakov:2004be}.

\begin{figure}[h!] 
\begin{picture}(100,105)
\put(-10,-7){\includegraphics{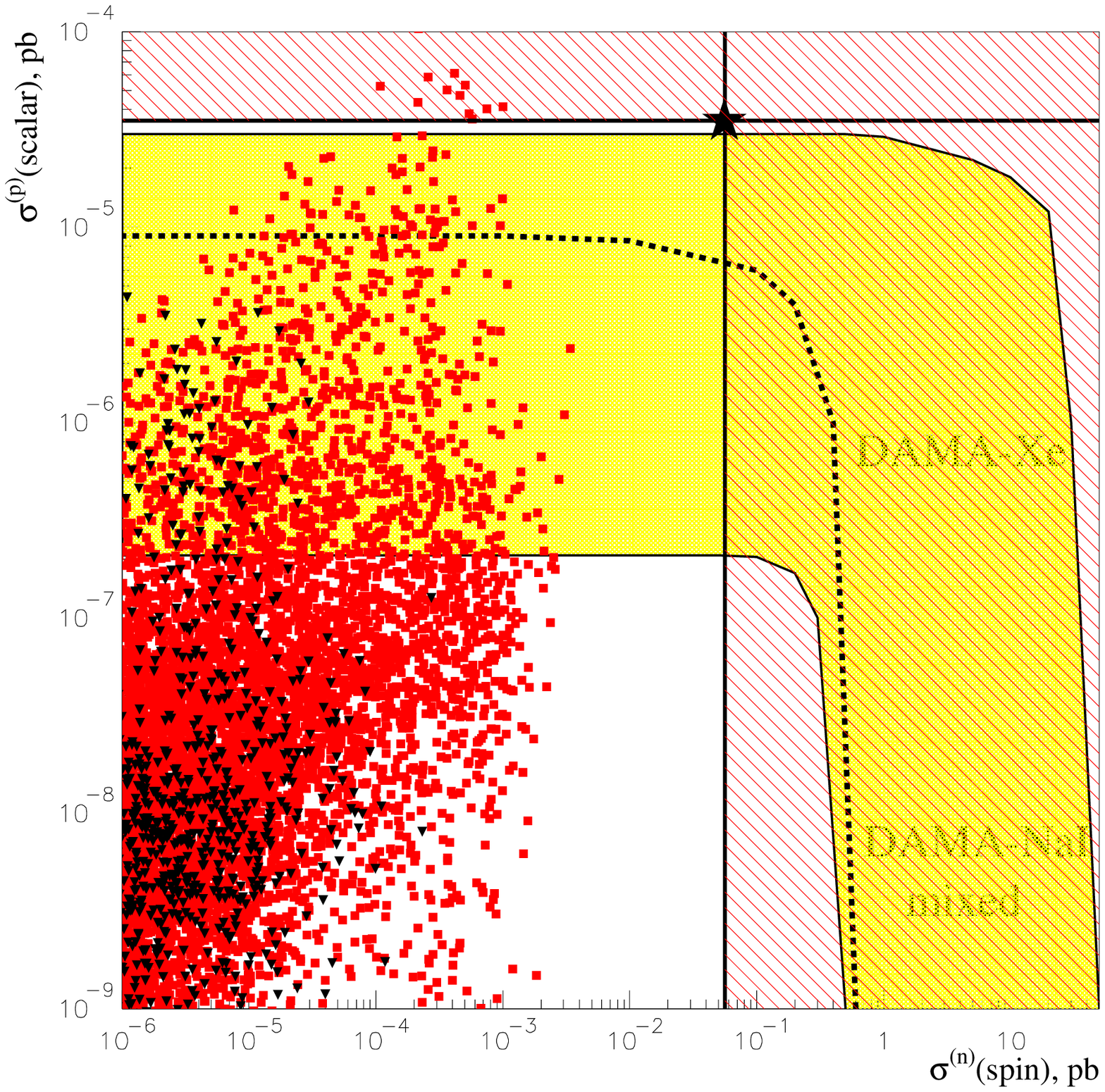}}
\end{picture}
\caption{
        The DAMA-NaI allowed region (inside yellow band) 
        for         SD WIMP-neutron coupling 
        versus SI WIMP-nucleon coupling is from 
\cite{Bernabei:2003za} and corresponds to 
        $40~{\rm GeV}/c^2 < m^{}_{\chi} < 110~{\rm GeV}/c^2$.
        The scatter plots 
from~\cite{Bednyakov:2004be}
        give correlations
        between $\sigma^{p}_{{\rm SI}}$ and 
        $\sigma^{n}_{{\rm SD}}$ in the effMSSM 
        for $m_\chi<200$~GeV.
        The squares (red) correspond to sub-dominant relic neutralino
        contribution $0.002 < \Omega_\chi h^2_0<0.1$    
        and triangles (black)
        correspond to WMAP relic neutralino density 
        $0.094 < \Omega_\chi h^2_0<0.129$. 
        The dashed line from 
\cite{Bernabei:2001ve}
        shows the DAMA-LiXe (1998) exclusion curve 
        for $m^{}_{\chi}=50$~GeV$/c^2$.
        The star gives our simultaneous upper limits 
	from the HDMS experiment 
\cite{Bednyakov:2007yf}
	for 
        $\sigma^{}_\SI$ and $\sigma^{n}_\SD$ for 
        $m^{}_{\chi} = 80~{\rm GeV}/c^2$. 
        Therefore values of
        $\sigma^{}_\SI$ above the horizontal line and
        of $\sigma^{n}_\SD$ located right from the vertical line are
        excluded by our analysis.
}
\label{dama-mixed-vs-hdms}
\end{figure}

      In 
Fig.~\ref{dama-mixed-vs-hdms}
      the new HDMS-2006 limits 
      on the SD and SI WIMP-nucleon interactions
      are compared 
      with the relevant DAMA constraints extracted 
      from measurement of the
      annual signal modulation with a NaI target 
\cite{Bernabei:2003za} as well as with  
     calculation in the low-energy effective MSSM 
\cite{Bednyakov:2004be}.
         To perform the comparison with the DAMA allowed region 
         of $\sigma^{}_\SI$ and $\sigma^{n}_\SD$  
(Fig.~\ref{dama-mixed-vs-hdms})
	 only one WIMP mass  
         $m^{}_{\chi} = 80~{\rm GeV}/c^2$ 
         (the star in the figure)
         was chosen for illustration.
         The upper limits for all other WIMP masses, 
         $40~{\rm GeV}/c^2 < m^{}_{\chi} < 110~{\rm GeV}/c^2$,
         are {\em very close}\/
         to this point. 
        The point marked with the star gives our 
        simultaneous upper limits for 
        $\sigma^{}_\SI$ and $\sigma^{n}_\SD$ for 
        $m^{}_{\chi} = 80~{\rm GeV}/c^2$. 
        For this WIMP mass, 
        values of $\sigma^{}_\SI$ located above the horizontal line and
        of $\sigma^{n}_\SD$ located to the right side 
        of the vertical line are excluded.
        Therefore, for any fixed WIMP mass in the domain   
        $40~{\rm GeV}/c^2 < m^{}_{\chi} < 110~{\rm GeV}/c^2$
        SUSY-like calculations 
        should give simultaneously 
        $\sigma^{}_\SI$ below the relevant (to the fixed WIMP mass) 
        horizontal line and 
        $\sigma^{n}_\SD$ to the left side of the relevant 
        vertical line, respectively.
        These limits improve the DAMA-Xe limit significantly 
	(about one order of magnitude)
	and exclude a 
        DAMA allowed region of large 
	spin-dependent WIMP-neutron cross sections.

        Here, perhaps, is the right place to make another 
        remark concerning the possibility to compare results
        from DM search 
	experiments 
        with passive background reduction
	(like DAMA, HDMS, etc)
        and 
	experiments with (mostly)  
        active background reduction (like CDMS, EDELWEISS, etc).
        First, we note, that obviously 
        any extra positively defined background-like contribution 
        to the spectra will decrease the extracted 
        (upper limit) values of the SD and SI cross section.
        Next, within the passive background reduction scheme 
        the measured spectrum is not affected by 
        hard- or software influence during the data taking.
        Extra 
        further background reduction can be done off-line 
        on the basis of careful investigation of 
        the spectrum itself or, 
        for example, with help of pulse shape analysis.
        In this case the extracted background contribution
        is under control and well defined.
        On the other side, within the active background reduction
        approach the 
        measured spectrum already contains results
        of this active reduction influence on 
        the data taking process.
        In this case it is   
        not simple to hold under control the real level 
        of extracted on-line background contribution
        which easily can be overestimated
(see, for example, the recent discussion of the ZEPLIN-I sensitivity in  
\cite{Benoit:2005gg}). 
        Therefore, due to this obvious difference
        a direct comparison of exclusion curves from 
        experiments with passive and active background
        reductions could be, in principle, rather misleading.

\section{Discussions} 
	The problem of the dark matter in the Universe
	is a challenge for modern physics and 
	experimental technology.
	To solve the problem, i.e. {\em at least}\/ 
	to detect dark matter particles, 
	one simultaneously needs to apply 	the front-end knowledge 
	of modern Particle Physics, Astrophysics,  
	Cosmology and Nuclear Physics as well as 
	one should develop, and use
	over long time 
	extremely high-sensitive experimental setups, and
	complex data analysis methods.

       Weakly interacting massive particles (WIMPs) 
       nowadays are among the best motivated 
       non-baryonic dark matter candidates. 
       In particular, 
       the lightest neutral supersymmetric particle (LSP), 
       the neutralino, is a very good WIMP candidate. 
       The motivation for supersymmetry arises
       naturally in modern theories of particle physics. 

       To estimate the expected direct detection rate for these WIMPs
       any SUSY-like model, for example, 
       an effective low-energy minimal supersymmetric extension 
       of the Standard Model (effMSSM), 
       or some measured data, for example, from the DAMA experiment 
\cite{Bednyakov:2005qd},  
       can be used. 
       On this basis the WIMP-proton and WIMP-neutron
       spin and scalar cross sections at zero-momentum transfer
       ($\sigma^{p,n}_{\SD}(0)$ and $\sigma^{p,n}_{\SI}(0)$)
       can be  calculated. 
       These calculations one usually compares with
       measurements, which (with the only exception of the DAMA result)
       are presented in the form of 
       exclusion curves --- upper limits of cross section 
       as functions of the WIMP mass.
       In the case of non-observation of any DM signal 
       an exclusion curve simply reflects the sensitivity 
       of a given direct DM search experiment and
       potentially 
       allows one to constrain some version of SUSY-like
       theory, {\em if the curve is sensitive enough}.
       Therefore the best exclusion curve is currently 
       a clear aim of almost all  
       dark matter search experiments (DAMA, LIBRA, and 
       GENIUS perhaps are/were the only exceptions).
       The main competition between the experiments 
       runs in the field of these exclusion curves.

       Before 2000  all exclusion curves
       were evaluated mainly in the one-coupling dominance approach
       (when only one cross section was defined from the measured spectra
       for fixed WIMP mass), 
       which gave slightly pessimistic 
       (for spin-non-zero target
       experiments), but universal limits for all experiments.
       One would say that the competition between DM experiments
       was honest. The predictions from SUSY-like models were 
       in general far  from being reached by the data. 
       
       Mainly after the paper
\cite{Tovey:2000mm} was published in 2000 
       (and as well after the DAMA evidence
\cite{Bernabei:2003za}) a new kind of exclusion curves appeared. 
	In particular, for the first time these curves were 
	obtained for the spin-dependent WIMP-nucleon cross section
	limits when non-zero sub-dominant spin WIMP-nucleon 
	contributions were taken into account
\cite{Ahmed:2003su,Miuchi:2002zp}.
        This procedure obviously improved the quality of
	the exclusion curves.
        Therefore a direct comparison of an old-fashioned exclusion curve 
        with a new one could in principle mislead one to a wrong
	conclusion about better sensitivity of the
	more recent experiments. 
        There is generally some possible incorrectness 
        in the direct comparison of the exclusion curves 
        for the WIMP-proton(neutron) spin-dependent cross section 
        obtained with and without the
        non-zero WIMP-neutron(proton) spin-dependent contribution.
        Furthermore the above-mentioned incorrectness concerns 
        to a great extent the direct comparison
        of spin-dependent exclusion curves obtained with and without non-zero
        spin-independent contributions
\cite{Bernabei:2003za,Bernabei:2003wy}.
        Taking into account both 
        spin couplings $a_p$ and $a_n$ but ignoring 
        the scalar coupling $c_0$, one can easily arrive at  
        a misleading conclusion 
        especially for not very light target nuclei
        when it is not obvious that 
        (both) spin couplings dominate over the scalar one.
        To be consistent,  one has 
        to use a mixed spin-scalar coupling approach
        as for the first time proposed by the DAMA collaboration
\cite{Bernabei:2000qi,Bernabei:2003za,Bernabei:2003wy}. 

	It was argued in 
\cite{Bernabei:2000qi,Bernabei:2003za,Bernabei:2003wy,%
Bednyakov:2005qp,Bednyakov:2004be}
	that
        potentially misleading discrepancies between the results of 
        different dark matter search experiments 
        (for example, DAMA vs CDMS and EDELWEISS) 
        as well as between the data and the SUSY calculations  
        can be avoided by using the 
        mixed spin-scalar coupling approach, where the 
        spin-independent and spin-dependent 
        WIMP-nucleon couplings are a priori considered to be
        {\em both}\/ non-zero.  

        The mixed spin-scalar coupling approach 
        was applied to
        analyze the data from both HDMS experiments 
        with natural Ge and with 
        the neutron-odd group high-spin isotope $^{73}$Ge. 
        The approach allows both upper limits for 
        spin-dependent
        $\sigma^{n(p)}_{{\rm SD}}$ 
        and spin-independent 
        $\sigma^{}_{{\rm SI}}$
        cross sections of 
        WIMP-nucleon interaction to be simultaneously 
        determined from the experimental data.
        In this way visible improvement in form of 
        exclusion curves is achieved relative to
        the traditional one-coupling dominance scheme
\cite{Bednyakov:2007yf}.
        The agreement of the obtained
        $\sigma^{n}_{{\rm SD}}$ and
        $\sigma^{}_{{\rm SI}}$ 
        with parameter regions allowed
        from the observation of the annual modulation signature
        by the DAMA collaboration is demonstrated. 
        The above-mentioned correlations between 
        $\sigma^{n}_{{\rm SD}}$ and
        $\sigma^{}_{{\rm SI}}$ 
        can be considered as
        a new requirement, which demands that
        for any fixed WIMP mass $m_\chi$ one should have \
        $\sigma^{}_\SI({\rm theor.})
        \le \sigma^{}_\SI({\rm fitted})$ 
        \ and \
        $\sigma^{n}_\SD({\rm theor.})
        \le \sigma^{n}_\SD({\rm fitted})$,
        simultaneously, 
        provided 
        $\sigma^{n(p)}_{\SD(\SI)}({\rm theor.})$
        are 
        calculated in any underlying SUSY-like 
        theory.
	For the first time a similar result for 
	NaI was mentioned by the DAMA collaboration
\cite{Bernabei:2001ve}. 

\enlargethispage{\baselineskip}
	It is important to note, 
	that without proper knowledge of the
	nuclear and nucleon structure it is not possible to 
	extract reliable and useful information (at least in form of these 
	$\sigma^{n}_{{\rm SD}}$ and
        $\sigma^{}_{{\rm SI}}$ cross sections) 
	from direct dark matter search experiments.
	However,  astrophysical uncertainties, in particular the DM 
	distribution in vicinity of the Earth
\cite{Copi:2002hm,Kurylov:2003ra,Tucker-Smith:2004jv,Gelmini:2004gm,%
Savage:2004fn,Gondolo:2005hh,Gelmini:2005fb}, make the
	problem of interpretation of results of 
	the DM search experiments far more complicated.
	At the moment to have a chance to compare sensitivities
	of different experiments people adopted one
	common truncated Maxwellian DM particle distribution, 
	but nobody can prove its correctness.
	Only in the case of indeed direct DM detection
	one can make some conclusions about the real
	DM particle distribution in the vicinity of the Earth.

       Furthermore, almost by definition (from the very beginning), 
       a modern experiment aiming at the best exclusion curve 
       is doomed to non-observation of the DM signal. 
       This is due to the fact, that a typical expected DM-signal 
       spectrum exponentially drops
       with recoil energy  and        it is practically  
       impossible to single it out from a background non-WIMP
       spectrum of a typical (semiconductor) detector, 
       which is as usual exponential as well.

       In fact, one needs some clear, or ``positive'' signature
       of WIMP particles interactions with target nuclei.
       Only exclusion curves are not enough.
       Ideally this signature should be a unique
       feature of such an interaction
(see for example 
\cite{Spooner:2007zh}).

       There are some typical characteristics 
       of WIMP particle interactions with a nuclear target which 
       can potentially play the role of such positive WIMP signatures
       (see for example
\cite{Gascon:2005xx}). 
   First of all WIMPs produce nuclear recoils, 
   while most radioactive backgrounds produce electron recoils. 
   Nevertheless, for example, neutrons 
   (and any other heavy neutral particle) also can produce nuclear recoils. 
   There exist also some proposals 
   which rely on WIMP detection via electron recoils 
(see for example \cite{Vergados:2005as,Vergados:2004qj,Vergados:2003st}).

   Due to the extremely rare event rate of the WIMP-nuclear interactions
   (the mean free path of a WIMP in matter is of the order of a light-year)
   one can expect two features.
   The first one is that
   the probability of two
   consecutive interactions in a single detector or two 
   closely located detectors is completely negligible. 
   Multiple interactions of photons, gamma-rays or neutrons under the
   same conditions are much more common. 
   Therefore only non-multiple interaction events 
   can pretend to be from WIMPs.
   The second one is a uniform distribution of the 
   WIMP induced events throughout a detector.
   This feature can also be used 
{\em in future}\/ 
   to identify background 
   events (from photons, neutrons, beta and alpha particles) 
   in rather large-volume position-sensitive detectors.

    The shape of the WIMP-induced recoil energy spectrum
    can be predicted rather accurately 
    (for given WIMP mass, for 
    fixed nuclear structure functions and astrophysical parameters). 
    The observed energy spectrum, pretending to be from WIMPs, 
    must be consistent with the expectation. 
    However, this shape is exponential, right as it is the case for many
    background sources.

   Obviously, 
   the nuclear-recoil feature, 
   the non-multiple interaction,
   the uniform event distribution throughout a detector
   and the shape of the recoil energy spectrum
   could not be a clear ``positive signature''
   of WIMP interactions.  
   One believes that the following three features of WIMP-nuclear 
   interaction can serve as a clear ``positive signature''.

   The currently most promising, technically reachable 
   and already used (by the DAMA collaboration) 
``positive signature'' is
   the annual modulation signature. 
   The WIMP flux and its average kinetic energy vary annually 
   due to the combined motions of the Earth and Sun 
   relative to the galactic center.
   The impact WIMP energy increases (decreases) when 
   the Earth velocity is added to (subtracted from) the velocity 
   of the Sun.
      The amplitude of the annual modulation depends on 
      many factors --- 
      the details of the halo model, mass of the WIMP, 
      the year-averaged rate (or total WIMP-nuclear cross sections), 
      etc.
      In general the expected modulation amplitude is rather small
(see for example, 
\cite{Freese:1987wu,Lewin:1996rx} and
\cite{Bernabei:2003za,Bernabei:2003wy}) 
      and to observe it one needs huge (at best ton scale)  detectors 
      which can continuously operate over 5--7 years.
      Of course, to reliably use this signature one should prove 
      the absence of annually-modulated backgrounds.
      One should, however, also be aware that seasonal modulation
      can also originate from other scenarios such as caustic rings
      of axions or neutralinos in the halo dark matter distribution
\cite{Sikivie:1999vv,Sikivie:2003uu}.

      Another potentially promising positive WIMP signature 
      is connected with the possibility to measure the 
      direction of the recoil nuclei induced by a WIMP.
      In these directional recoil experiments 
      one plans to measure the correlation of the event 
      rate with the Sun's motion
(see for example, \cite{Vergados:2003pk,Vergados:2002bb,Vergados:2004qj}).
      Unfortunately, the task is extremely complicated 
(see for example,
\cite{Morgan:2004ys,Sekiya:2004ma,Alner:2004cw,%
Snowden-Ifft:1999hz,Gaitskell:1996cv}).

        The third well-known potentially useful positive WIMP signature 
	is connected with the coherence of the WIMP-nucleus 
	spin-independent interaction. 
	Due to a rather low momentum transfer a WIMP 
	coherently scatters on the whole target nucleus and 
	the elastic cross-section of this interaction 
	should be proportional to $A^{2}$, where $A$ is 
	the atomic number of the target nucleus.
	Contrary to the $A^2$-behavior, 
        the cross-section of 
	neutron scattering on nuclei 
	(due to the strong nature of this interaction)
	is proportional 
	to the geometrical cross-section of the target nucleus 
	($A^{2/3}$-dependence). 
	To reliably use this $A^2$-signature one has to satisfy
	at least two conditions.
	First, one should be sure that the spin-independent 
	WIMP-nuclear interaction indeed dominates over
	the relevant spin-dependent interaction. 
	This is far from being obvious
(see for example, 
\cite{Bednyakov:2007zz,Bednyakov:2004be,Bednyakov:2003wf,%
Vergados:2005ky,Vergados:2004hw}).
         Second, one should, at least, 
	 for two targets with different atomic number $A$   
	 rather accurately 
	 measure the recoil spectra 
	 (in the worst case integrated event rates) 
	 under the same background conditions.
	 Currently this goal looks far from being realizable.

	 Developing further the idea of this third signature, 
	 one can also consider as a possible extra WIMP-signature
	 an observation of 	 the similarity
	 (or coherent behavior)
	 of measured spectra at different (also non-zero spin)
	 nuclear targets.
	 This possibility relies on rather accurate 
	 spin structure functions 
	 for the experimentally interesting nuclei
(see for example, 
\cite{Bednyakov:2006ux,Bednyakov:2004xq}). 

       Also in the case of currently very promising event-by-event
       active background reduction techniques (like in 
       the CDMS and EDELWEISS experiments)
       one inevitably needs clear positive WIMP signature(s).
       Without these signatures  
       one hardly can convince anyone that 
       the final spectrum is saturated only by WIMPs.
       Furthermore with the help of these extra signatures  
       one can define the WIMP mass from the spectrum
\cite{Green:2007rb,Shan:2007vn}.

        It is known (see for example the discussion in 
\cite{Gondolo:2005qp} and earlier partly in  
\cite{Bednyakov:1999vh}) that a proof of
        the observation of a dark matter signal
        is an extremely complicated problem.
	As pointed out above, 
        on this way an interpretation of measurements  
        in the form of exclusion curves helps almost nothing. 
        Of course, an exclusion curve 
        is at least something from nothing observed.
        It allows sensitivity comparison
        of different experiments and therefore allows  
        to decide who at the moment is the best 'excluder'.
        But, for example, supersymmetric theory is, 
        in general, very flexible, it has a lot of parameters, 
	and one hardly believes   that an exclusion curve 
	can ever impose any decisive constraint on it. 
        The situation is much worse due to the
	already mentioned 
        famous nuclear and astrophysical uncertainties involved
        in the exclusion curves evaluation
\cite{Kinkhabwala:1998zj,Donato:1998pc,Evans:2000gr,%
Green:2000jg,Green:2001xy,Copi:2000tv,Ullio:2000bf,Vergados:2000cp}.
        This is why, from our point of view, it is not very 
        decisive (or wise) to use very refined 
        data and methods (nuclear, astrophysical, numerical, statistical 
\cite{Feldman:1997qc}, etc)
        and spend big resources fighting 
        only for the best exclusion curve.
	This fighting could be only accepted, perhaps, in the case
	when one tries to strongly improve the sensitivity of
	a small detector having future plans to use many copies
	of it in a huge detector array with a total 
	ton-scale mass.  

        As it already has been stressed in 
\cite{Klapdor-Kleingrothaus:2005rn,Baudis:2000ph}, 
        in case of a positive DM signal, e.g. the detector 
	HDMS has no means 
        to discriminate the signal from background.
        With a target mass of 200 g only,        the statistical
        accuracy within 2 years of measurements is 
        too low in order to see the annual modulation, which 
        is nowadays the only available positive 
        signature of WIMP interaction with terrestrial matter. 
        The same is completely true for 
        any other potentially very accurate 
	low-target-mass direct dark matter search experiment.
        To have a chance to see the
        annual modulation signature
        of WIMP-nuclear
        interaction and to detect dark matter particles, 
	as seems to have been done by DAMA,
        one 
	preferably 
	needs either a GENIUS-like huge setup
\cite{Klapdor-Kleingrothaus:1997pw,Hellmig:1998pv}
        which was planned to operate up to 1000 kg of HPGe
        detectors 
of different enrichment of $^{76}$Ge and $^{73}$Ge  
        (in a large volume of ultra-pure liquid nitrogen), 
        or, perhaps, a setup with a bit smaller mass, which 
        is able to perform permanent data taking over at least several 
        years under extremely low background conditions
(like for example, the GENIUS-TF experiment 
\cite{Tomei:2003vc,Klapdor-Kleingrothaus:2000ke}, or 
     a future enlarged EDELWEISS setup). 
     The performing
     of such experiment seems, however, more difficult than
     originally expected
\cite{Klapdor-Kleingrothaus:2003pg,KlapdorKleingrothaus:2004ev,%
KlapdorKleingrothaus:2006rf,Krivosheina:2006zv}.

\section{Conclusion} 
        In this review paper the following main questions 
	have been discussed.

{\em 	Why do we want to improve the exclusion curves?}\/
	The answer usually is:
	to constrain a SUSY-like theory. 
	Unfortunately this is an 
	almost hopeless aim due to the huge flexibility of 
	such theories and the inevitable necessity of extra
	information from other SUSY-sensitive observables 
	(for example, from LHC, or Tevatron).
	Almost all experimental groups presenting their 
	exclusion curves 	try to compare them with 
	some SUSY predictions. 
	It is clear from this comparison (see for example, 
Fig.~\ref{Scalar-2004}) 
	that there are some domains of the SUSY parameter
	space, which are excluded already now by these exclusion curves.
	What is remarkable, however, is 
	that nobody yet has seriously considered 
	--- or used otherwise --- these
	constraints for SUSY. 	
         In short, at the present and foreseeable level of 
	 experimental accuracy, simple fighting for the best exclusion curve 
	 is almost useless, either for real DM detection, or
	 for substantial restrictions for SUSY.

{\em How far one can improve an exclusion curve?}\/ 
	 It is almost a question of taste, 
	 when one should decide to stop speculations
         on the improvement of the exclusion curve. 
	 Almost always one can find something to improve 
	 the exclusion curve.

{\em What one would like to see in the future beyond an exclusion curve?}\/
     New generations of dark matter experiments 
     right from their beginning {\em should aim at 
     detection}\/ of dark matter particles.
     This will require development of new setups, 
     which will be able to register 
     {\em positive signatures} of the dark matter particles
     interactions with nuclear targets.
     At least the DAMA 
\cite{Bernabei:2000qi,Bernabei:2001ve,Bernabei:2003wy,Bernabei:2003za}
and LIBRA
\cite{Bernabei:2006kc,Bernabei:2006tz} 
       experiments are seen on the way.
    In order to be convincing, an eventual WIMP signal should combine
    more than one of these positive WIMP signatures
\cite{Gascon:2005xx,Spooner:2007zh}. 

{\em Why one should try to obtain a real recoil energy spectrum?}\/
        The spectrum allows one to look for the annual modulation effect, 
	the only nowadays available positive dark matter 
	signature, which can prove existence of dark matter 
	particle interactions with terrestrial nuclei.
	There are also attempts to determine the WIMP mass
	on the basis of measured recoil spectra
\cite{Green:2007rb,Shan:2007vn}.
	Very accurate off-line investigation of the measured spectrum 
	allows one to single out different non-WIMP background
	sources and to perform controllable 
	background subtractions.  

\enlargethispage{\baselineskip}
     It seems that, at the level of our present knowledge 
     the dark matter problem could not be solved independently 
     from other related problems (proof of SUSY, 
     astrophysical dark matter properties, etc).
     Furthermore, 
     due to the huge complexity (technical, physical, astrophysical, 
     necessity for positive signatures, etc)   
     to solve the problem of dark matter 
     one should not be afraid, but openly 
     use a reliable model-dependent framework --- 
     for example the framework of SUSY, where the same LSP neutralino 
     should be seen coherently or lead to effects
     in all available experiments 
     (direct and indirect DM searches, 
     rare decays, high-energy searches at LHC, etc).
     Only if such SUSY framework leads to a specific and decisive 
     positive WIMP signature, this could mean a proof of SUSY and 
     simultaneous solution of the dark matter problem.
     It is on the other hand absolutely clear, that SUSY 
     although in contrast to others being prefered, since requested by 
     'higher' particle physics theories, such as Superstrings, is not 
     the only candidate for the origin of dark matter, 
     and also other scenarios
     have to  be investigated in a comparably thorough way.
%

       This work was supported by the RFBR
       (grant 06--02--04003) and DFG (grant 436 RUS 113/679/0-2(R)).
       The authors thank Dr.~I.V.~Krivosheina 
       for the long-term and very fruitful collaboration, 
       and Dr.~E.A.~Yakushev (JINR) for useful discussions.

{\small 
\providecommand{\href}[2]{#2}\begingroup\raggedright\endgroup
 } 
\end{document}